\documentclass[useAMS,usenatbib]{mn2e}

\usepackage{graphicx,epsfig,upgreek,subfigure}
\def\lesssim{\!\!\!\phantom{\le}\smash{\buildrel{}\over{\lower2.5dd\hbox{$\buildrel{\lower2dd\hbox{$\displaystyle<$}}\over\sim$}}}\,\,}

\newcommand{\msol}{\mbox{M$_{\odot}$}}

\newcommand{\kms}{\mbox{$\rm{km}\,s^{-1}$}}

\def \aj {AJ}
\def \mnras {MNRAS}
\def \apj {ApJ}
\def \apjl {ApJL}
\def \aap {A\&A}

\def \araa {ARAA}

\def \pasp {PASP}

\def \apjs {ApJS}
\def \aaps {A\&AS}

\title{Supernovae and radio transients in M~82}
\author[Mattila et al.]
{S. Mattila\thanks{seppo.mattila@utu.fi}$^{1,2}$, M. Fraser$^{3}$, S.J. Smartt$^{3}$, W.P.S. Meikle$^{4}$, C. Romero-Ca\~nizales$^{2}$,
\newauthor R.M. Crockett$^{3,5}$, A. Stephens$^{6}$ \\
$^{1}$Finnish Centre for Astronomy with ESO (FINCA), University of Turku, V\"ais\"al\"antie 20, FI-21500 Piikki\"o, Finland.\\
$^{2}$Tuorla Observatory, Department of Physics and Astronomy, University of Turku, V\"ais\"al\"antie 20, FI-21500 Piikki\"o, Finland.\\
$^{3}$Astrophysics Research Centre, School of Mathematics and Physics, Queen's University of Belfast, Belfast BT7 1NN, UK.\\
$^{4}$Astrophysics Group, Blackett Laboratory, Imperial College London, Prince Consort Road, London, SW7 2AZ, UK.\\
$^{5}$Oxford Astrophysics, Department of Physics, Denys Wilkinson Building, Keble Road, Oxford, OX1 3RH, UK.\\
$^{6}$Gemini Observatory, 670 North Aohoku Place, Hilo, HI 96720, USA.\\
}

\date{}

\begin{document}
\maketitle
\begin{abstract}
We present optical and near-infrared (IR) photometry and near-IR spectroscopy
of SN~2004am, the only optically detected supernova (SN) in M~82. These
demonstrate that SN~2004am was a highly reddened type II-P SN similar to the
low luminosity type II-P events such as SNe 1997D and 2005cs. We show that SN~2004am was
located coincident with the obscured super star cluster M~82-L, and from
the cluster age infer a progenitor mass of $12^{+7}_{-3}$\,\msol. In addition to
this, we present a high spatial-resolution Gemini-N $K$-band adaptive optics image of the site of 
SN~2008iz and a second transient of uncertain nature, both detected so 
far only at radio wavelengths. Using image subtraction techniques together 
with archival data from the {\it Hubble Space Telescope}, we are able to 
recover a near-IR transient source co-incident with both objects. We find
the likely extinction towards SN~2008iz to be not more than A$_{\rm V}$ $\sim$ 10.
The nature of the second transient remains elusive and we regard an extremely
bright microquasar in M~82 as the most plausible scenario.
\end{abstract}


\section{Introduction}
In the nuclear regions of M~82 and other nearby starburst galaxies one 
core-collapse supernova (CCSN) is expected to explode every 5-10 years and
a number of young supernova remnants (SNRs) have been revealed
in these regions by radio observations (\citeauthor{1985ApJ...291..693K 1985).}
However, due to the high dust extinction in the nuclear starburst regions most 
of the actual SNe have remained undetected (e.g., \citeauthor{2001MNRAS.324..325M} 2001). 
Searches working at near-infrared (NIR) wavelengths have discovered SNe in the
nuclear and circumnuclear regions of a few nearby starburst galaxies and luminous infrared galaxies (LIRGs) 
(e.g., \citeauthor{2012ApJ...744L..19K} 2012; \citeauthor{2008ApJ...689L..97K} 2008; \citeauthor{2010CBET.2446....1M} 2010;
\citeauthor{2007ApJ...659L...9M} 2007; \citeauthor{2002A&A...389...84M} 2002).
Furthermore, optical searches have been able to discover a few SNe in normal nearby spiral galaxies with several 
magnitudes of visual extinction (e.g., SNe 2002hh, 2002cv; \citeauthor{2006MNRAS.368.1169P} 2006;
\citeauthor{2002A&A...393L..21D} 2002).

The first claim of a detection of a supernova at optical or NIR 
wavelengths in M~82 was by \citeauthor{1986IAUC.4197....2L} (1986), who reported the 
discovery of SN~1986D in 2 $\upmu$m observations. However, it was later shown 
by \citeauthor{1986IAUC.4202....2G} (1986) that the source identified as SN~1986D was 
already present in their 2.2 $\upmu$m  images obtained three years earlier and 
has not varied significantly in brightness since then. This source was 
designated as the feature K2 by \citeauthor{1986AJ.....91..758D} (1986). 
SN~2004am in M~82 was discovered (\citeauthor{2004IAUC.8297....2S} 2004) by the Lick Observatory SN Search (LOSS)
on images from 2004 Mar. 5.2 UT but it was already present in their earlier images from 2003 Nov. 21.6.
Being on a bright spot in M~82 caused their detection software to initially miss the new object.
On 2004 March 6.9 SN~2004am was spectroscopically classified as a highly reddened type II event
showing some spectral similarity to the type II-P SN~1995V (\citeauthor{2004IAUC.8299....2M} 2004). Radio
non-detections of SN~2004am on 2003 Nov. 14 (8.4 and 15 GHz) and on 2004 March 9 (5 GHz) were reported
by \citeauthor{2004IAUC.8332....2B} (2004). SN~2004am is still the only SN ever {\it discovered} at optical/IR
wavelengths in M~82.

Being a prototypical starburst galaxy at a distance of $3.3\pm0.3$ Mpc, (from the Cepheid distance to the
M~81 group; \citeauthor{2001ApJ...553...47F} 2001) means that there is a wealth of ground and space based imaging of M~82.
Identification of progenitors in both {\it Hubble Space Telescope} ({\it HST}) and ground-based images has now 
become frequent (e.g., \citeauthor{2004Sci...303..499S} 2004; \citeauthor{2006ApJ...641.1060L} 2006;
\citeauthor{2008ApJ...688L..91M} 2008; \citeauthor{2009Sci...324..486M} 2009; \citeauthor{Fra11} 2011; \citeauthor{Van12} 2012).
Furthermore, in M 82 the compact star population has been studied extensively (e.g., \citeauthor{2008A&A...486..165L} 2008;
\citeauthor{2007MNRAS.379.1333B} 2007; \citeauthor{2006MNRAS.370..513S} 2006; \citeauthor{2003ApJ...596..240M} 2003). Hence
SN~2004am is interesting not just because it is the first SN discovered in M~82 in the optical and NIR, but also because
its physical association with the well studied super star cluster (SSC) M~82-L allows a mass estimate of the progenitor star.
In one recent case, SN~2004dj in NGC2403, the explosion was coincident with a young compact star cluster
(\citeauthor{2004ApJ...615L.113M}, 2004; \citeauthor{2005ApJ...626L..89W} 2005). The spectral energy distributions of the
host cluster suggest ages and masses 13-20~Myr and 2.4-9.6$\times 10^4$~\msol. If we assume that the stellar population
in the cluster is coeval, then this leads to estimates of the progenitor mass of 12-15\msol. \citeauthor{2009ApJ...695..619V}
(2009) produced an improved study of the host cluster once SN~2004dj faded using extensive data from the UV to NIR.
They estimated a most likely turn-off mass of between 12 and 20~\msol. SN~2004dj was a type II-P
(\citeauthor{2006MNRAS.369.1780V} 2006), and the estimate of the progenitor mass from the turn-off age is consistent
with the mass estimates from the direct detection of SN II-P progenitors on pre-explosion images
(\citeauthor{2009MNRAS.395.1409S} 2009).

Over the last thirty years, there have been four detected radio transients in
M~82 which have been suggested to be SNe. The source
41.5$+$59.7 (\citeauthor{1985Sci...227...28K} 1985) was seen in February 1981 at 5\,GHz
with a flux density of 7.07$\pm$0.24\,mJy, but by April of the following year
had faded below the detection limit of 1.5\,mJy. Another transient 40.59$+$
55.8 (\citeauthor{1994MNRAS.266..455M} 1994) was detected at one epoch with the
Multi-Element Radio Linked Interferometer Network (MERLIN) at 5\,GHz with a
flux density of $\sim$1.23\,mJy in July 1992. Neither of these two sources
were detected by \citeauthor{2008MNRAS.391.1384F} (2008) to a flux limit of
$\sim$20\,$\mu$Jy suggesting them to be transient sources, as opposed to
sources with variable nature.

More recently, a bright ($\sim$100 mJy) radio transient in the nuclear regions
of M~82 was discovered by \citeauthor{2009AA...499L..17B} (2009) in 22\,GHz images from the
Very Large Array (VLA) obtained in March 2008. Over one year's time the source
faded by a factor of ten, showing a spectral index of about $-0.8$ in the VLA
(1.4-43\,GHz) observations from April 2009 indicating an optically thick
synchrotron spectrum. Furthermore, very long baseline interferometry (VLBI)
observations from May 2008 and April
2009 revealed a ring-like structure expanding by $\sim$23\,000\,km\,s$^{-1}$
(\citeauthor{2010A&A...516A..27B} 2010). These observations definitely confirmed the SN
nature of the transient designated as SN~2008iz. Monthly monitoring of M~82
with the 25\,m Urumqi radio telescope at 5\,GHz led to the detection of SN\,2008iz
as a flare on top of the M~82 radio emission (\citeauthor{2010A&A...509A..47M} 2010). The SN
peaked with a flux of $\sim160$\,mJy on 21st June 2008, while modelling of its light
curve yielded an accurate explosion date of 18th February 2008.

In May 2009, another new radio transient was discovered in the nuclear regions
of M~82 by means of MERLIN observations (\citeauthor{ATEL2073} (2009). 
Following Gendre et al. (2012) we call it the 43.78+59.3 transient.
This source reached a peak intensity of
0.72\,mJy at 5\,GHz in early May 2009, and 1.7\,mJy at 1.6\,GHz on 20th May
2009. The nature of this object is still unknown due to its peculiar low
luminosity, its longevity (still observed after nine months
of its discovery), its position with respect to the dynamical centre of the
galaxy, and tentative evidence of proper motion and expansion. At the moment,
the most viable explanation for its nature appears to be a micro-quasar
(\citeauthor{2010MNRAS.404L.109M} 2010; \citeauthor{2011MNRAS.415L..59J} 2011) displaying a high ratio of
radio to X-ray luminosity. For instance, the source Cygnus X-3, which is one of
the strongest Galactic micro-quasars, has a radio luminosity exceeding its
X-ray luminosity by an order of magnitude (\citeauthor{2005MNRAS.361..633N} 2005). Thus a
similar scenario could also be plausible in the case of the 43.78+59.3 transient.

In this paper we present photometric and spectroscopic observations of SN~2004am
which despite its small distance has remained unstudied due to the high
line-of-sight dust extinction and the fact it occured coincident with the nuclear
SSC M~82-L. We make use of this coincidence to infer an initial mass for the progenitor.
We use deep, high spatial-resolution imaging from Gemini-N
to search for NIR counterparts of SN~2008iz and the 43.78+59.3 transient
and make use of these to investigate their nature and the line-of-sight extinctions
towards these sources. Finally, we discuss the nature of the radio transients
in M~82 making use of existing radio observations and compare these with the
expectations from the SN rate estimates.

\begin{table*}
\begin{center}
\caption{Optical and NIR photometry of SN~2004am. The epochs are assuming the first observation of the SN by LOSS was 2 weeks after the explosion
(see Sect. 2.3). Between 14 and 118 days unfiltered CCD magnitudes are from Singer, Pugh \& Li (2004). The ``und.'' labels means the SN was
undetected at these wavelengths with any significance in comparison to the coincident host cluster.}
\label{naco_obs}
\begin{tabular}{llccccccr}
 \hline
Date (UT)  &  Epoch & $V$ & $unfilt$ & $I$ & $J$ & $H$ & $K$ & Instrument/Reference\\ 
(2004)     & (days)\\ \hline
Nov. 21.5  &  14     & -  & 16.0 & -    & -    & -    & -    & Singer, Pugh \& Li (2004)\\
Nov. 23.5  &  16     & -  & 16.1 & -    & -    & -    & -    & Singer, Pugh \& Li (2004)\\
Dec. 28.4  &  51    & -  & 16.3 & -    & -    & -    & -    & Singer, Pugh \& Li (2004)\\
Jan. 2.5   &  56    & -  & 16.4 & -    & -    & -    & -    & Singer, Pugh \& Li (2004)\\
Jan. 17.4  &  71    & -  & 16.4 & -    & -    & -    & -    & Singer, Pugh \& Li (2004)\\
Jan. 21.4  &  75    & -  & 16.4 & -    & -    & -    & -    & Singer, Pugh \& Li (2004)\\
Jan. 30.3  &  84    & -  & 16.4 & -    & -    & -    & -    & Singer, Pugh \& Li (2004)\\
Feb. 5.3   &  90    & -  & 16.4 & -    & -    & -    & -    & Singer, Pugh \& Li (2004)\\
Mar. 5.2   &  118     & -  & 17.0 & -    & -    & -    & -    & Singer, Pugh \& Li (2004)\\
Mar. 6.9   &  120     & - & - & - & 12.95$\pm$0.08 & 12.49$\pm$0.16 &  12.03$\pm$0.14 & WHT LIRIS\\
June 5.9   &  211     & - & - & - & - & - & 13.06$\pm$0.12 & WHT LIRIS\\
July 5.3   &  241     & und. & - & $>20.3$ & - & - & -  & HST ACS/HRC\\
Nov. 25.2  &  384     & -       & - & -       & und. & und. & und. & WHT LIRIS\\            
\hline
\end{tabular}
\end{center}
\end{table*}

\section{SN~2004am: Observations and results}
\subsection{Observations and data reduction}
SN~2004am was observed with the NIR imager and spectrograph
LIRIS on the William Herschel Telescope (WHT). LIRIS spectra
covering $zJ$ bands were obtained on two epochs, 6 March and 25
November 2004. The observations included in this study are listed
in Table 1. The NIR imaging data were reduced using standard {\sc IRAF} routines
and the spectroscopy using the {\sc FIGARO} package as a part of the
{\sc Starlink} software. SN~2004am appears coincident with the SSC M~82-L
in these images (with seeing FWHM $\sim$ 1.5''), and therefore,
all the photometry was performed on subtracted images. The ISIS2.2 package
(\citeauthor{1998ApJ...503..325A} 1998) was used to convolve the better seeing image to the
poorer seeing one, and to match the intensities and background prior to the image
subtraction. The 25 Nov. 2004 $JHK_{\rm s}$ images were used as the image subtraction
references for the LIRIS images, as they provided a smooth subtraction and a further
$K_{s}$ epoch on 30 January 2005 indicated that the SN was no longer detectable in the 25 Nov $K_{s}$ frame.

In addition, we recovered optical images containing the site of SN~2004am from the
Hubble Space Telescope (HST) archive. The pipeline-reduced products from the HST archive
were used. SN 2004am was observed with the Advanced Camera for Surveys (ACS) HRC instrument
in F555W and F814W filters on 5 July 2004 (SNAP 10272; PI: A. Filippenko). The site of
SN~2004am was covered also with the ACS WFC instrument in F814W filter on 9 Feb. 2004
(PID 9788; PI: L.C. Ho) and 27 March 2006. Unfortunately, SN~2004am was saturated in the
image from 9 Feb 2004 and therefore we could not make
use of it for reliable photometry. However, we were still able to use it for precise
relative astrometry of SN 2004am (see Fig. 1). We attempted to detect SN~2004am in the ACS/HRC images
using images obtained with the same instrument and filters on 7 June 2002 as reference
for the image subtraction with the ISIS2.2 package. However, no significant source
was detected at the location of M~82-L in the difference images and we conclude that
SN~2004am had already faded below the detection limit of HST by the 5 July 2004 epoch
of observation (see Fig. 2).

\begin{table*}
\caption{Reported coordinates of the transients in M~82 discussed in this paper. All
coordinates are J2000.}
\label{t_coords}
\begin{tabular}{lccr}
\hline
Transient			& RA				& Dec			& Reference \\
\hline
SN~2004am		& 9:55:46.61		& +69:40:38.1		& \citeauthor{2004IAUC.8297....2S} (2004)\\
SN~2008iz		& 9:55:51.55		& +69:40:45.792 	& \citeauthor{ATEL2020} (2009)\\
			& 9:55:51.55		& +69:40:45.788	& \citeauthor{ATEL2060} (2009)\\
43.78+59.3 transient 	& 9:55:52.5083 	& +69:40:45.420	& \citeauthor{ATEL2073} (2009)\\	
\hline
\end{tabular}
\end{table*}

\subsection{Photometry of SN~2004am}
Photometry of SN~2004am was performed in the subtracted LIRIS images
using the aperture photometry procedure in the {\sc Gaia} image analysis tool
(Draper 2004). The JHK magnitudes from 2MASS for three bright field stars covered in the LIRIS field of view
were used for the photometric calibration. The average of the zeropoint magnitudes obtained
was adopted for the calibration and their standard deviation as the photometric uncertainty.
The statistical uncertainty in the photometry as estimated by {\sc Gaia} was in all cases
negligible.

The residual noise from the flux of M~82-L leaves no obvious point source in the
subtracted HST/ACS HRC images from 5 July 2004 (see Fig. 2). Aperture photometry was performed on the residuals using {\sc Gaia}.
For this we used a 0.25'' radius aperture and a sky annulus between 1.5 and 2.0 $\times$ the
aperture radius. No suitable point sources were present within the ACS/HRC field of view and thus we
adopted an aperture correction from 0.25'' to infinite aperture from Sirianni et al. (2005).
Application of the appropriate Vegamag zeropoint yielded m(F814W) = 20.3, which we adopt as an upper
limit for the brightness of SN 2004am at this epoch.

\begin{figure*}
\begin{center}
\includegraphics[width=175mm, angle=0, clip] {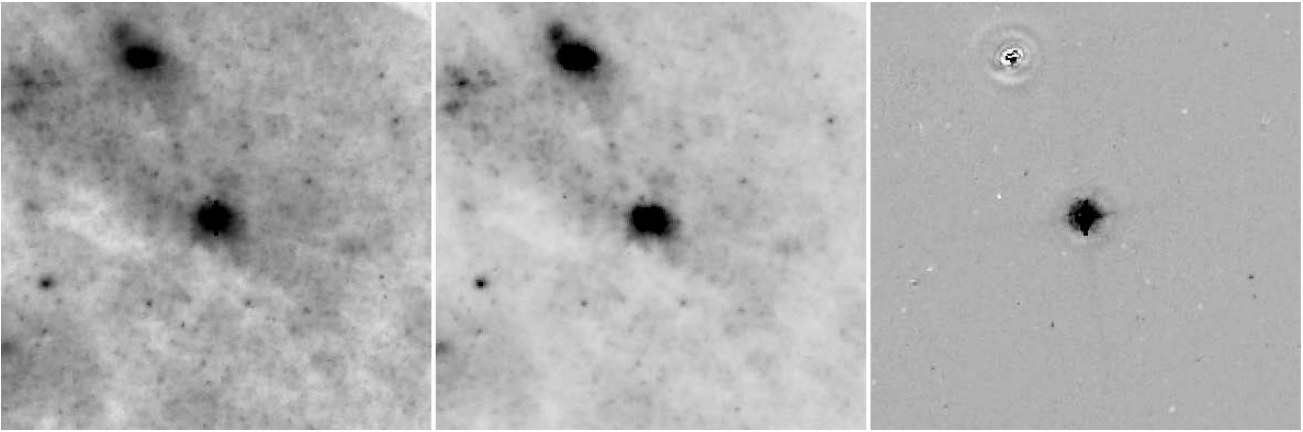}
\caption{10''$\times$10'' sections of ACS/F814W images (North is Up and East to the Left)
showing the SSCs M~82-L (in the middle) and M~82-F (to the north-east from the centre)
obtained on 9 February 2004 ({\bf left}) and 27 March 2006 ({\bf middle})
and the subtraction between the two obtained using ISIS2.2 ({\bf right}). The
images are shown with inverted intensity scale and the first two with log scaling.
A positive saturated image subtraction residual is visible in the subtracted image at
the location of M~82-L.}
\end{center}
\end{figure*}

\subsection{The type and extinction of SN~2004am}
The spectra of SN~2004am at both epochs are contaminated by the SSC. We 
subtracted the high signal-to-noise spectrum of M~82-L 
(provided by \citeauthor{2008A&A...486..165L} 2008) from both epochs. This spectrum was 
manually scaled and the simple subtraction procedure resulted in an apparently 
flat spectrum with the SSC continuum removed in the first epoch (6th March 
2004). However the subtraction from the second epoch left no identifiable SN 
features and we conclude that SN~2004am had faded below the detection limit of 
the spectral signal. 

In Figure 3, the subtracted SN~2004am spectrum is compared to the spectra of
three type II-P SNe: 2004et (Maguire et al. 2010), 1995V
(\citeauthor{1998MNRAS.299..150F} 1998) and 1997D (\citeauthor{2001MNRAS.322..361B} 2001). All spectra
were dereddened (by the values quoted in the original publications) and
scaled to the continuum level at 1.3$\mu$m of SN~2004am. SN~2004am was
dereddened with $A_{V} = 3.7$ and $R = 2.4$ as found for M~82-L by Lan{\c c}on
et al. (2008). SN~2004am has a similar spectrum to SNe II-P at the end of the
plateau phase. The ejecta velocity from the P-Cygni troughs in the metal lines
(e.g. Sr\,{\sc ii}, Fe\,{\sc ii}, and C\,{\sc i}) indicates that SN~2004am
appears more similar to the low velocity and low luminosity SN~1997D, than
the other two. Cross-correlation and fitting Gaussians to the centroids of the
sharpest P-Cygni profiles show that the velocity difference between SNe 2004am
and 1997D is negligible, while the line troughs of SNe 1995V and 2004et are
blueshifted by $1000\pm500$\kms. Furthermore the spectral features of
SNe 1995V and 2004et are visually broader (see Fig. 3). 

The classification of SN~2004am as a type II-P is supported by its unfiltered light
curve shown in Fig.\,3, compared with the $R$-band light curves of two different II-P events:
SNe 1999em and 2005cs (\citeauthor{2003MNRAS.338..939E} 2003; \citeauthor{2009MNRAS.394.2266P} 2009).
The plateau points of SN~2004am are unfiltered CCD magnitudes from Singer, Pugh \& Li (2004),
and the upper limit at 241 days has been estimated from the F814W HST images.
The light curves of the two supernovae have been scaled to the same
distance as SN~2004am, and have been dimmed to correspond to host galaxy extinctions
of A$_{R}$ = 5.4 and 4.2 mag, respectively. For this we adopted a distance of 11.7 Mpc
for the host galaxy of SN 1999em based on Cepheids (\citeauthor{2003ApJ...594..247L} 2003) and 8.4 Mpc
for the host galaxy of SN 2005cs based on the application of the Expanding Photosphere Method
for SNe 2005cs and 2011dh in the same host galaxy (\citeauthor{2012A&A...540A..93V} 2012). The total
(host galaxy + Galactic) extinctions of A$_{\rm V}$ = 0.31 and 0.43 were adopted for SNe 1999em
and 2005cs, respectively, following \citeauthor{2009MNRAS.395.1409S} (2009) and a Galactic extinction
of A$_{\rm V}$ = 0.53 (\citeauthor{1998ApJ...500..525S} 1998) towards M~82. Assuming an epoch of explosion of 14 days
before the SN was first observed by the LOSS was found to yield a good match with the post-plateau drop
in the light curves of SNe 1999em and 2005cs. If the extinction toward
SN~2004am is similar to that found for M~82-L by Lan{\c c}on et al. (2008) then its absolute
magnitude is fainter than that of the low-luminosity type II-P SN~2005cs.
SN~2004am is only likely to be a normal type II-P SN if the extinction is higher than
$A_{R}\simeq5$. Also, the spectrum analysis indicates that SN~2004am has a low ejecta velocity and
is similar to SNe 1997D and 2005cs. Furthermore, SN~2004am is not visible in the late deep
image from HST ACS at 241 days, allowing a conservative detection limit of
$I\simeq20.3$ to be set. The corresponding $R$-band limit would be fainter than this due to
the red $R-I$ colour of SNe II-P at such epochs and the likely high extinction towards SN 2004am.
Figure\,3 shows this is too faint to be compatible with
the nebular phase of SN 1999em, but is compatible with a SN that
ejected a low amount of $^{56}$Ni such as SN~2005cs (\citeauthor{2009MNRAS.394.2266P} 2009). 

Comparison of the unfiltered SN photometry from Singer, Pugh \& Li (2004) on 5 March
with our $K-$band photometry on 7 March yields a SN color of $(m_{\rm CCD} -K)\simeq +5.0$.
Typical $R-K$ magnitudes around 100 days for II-P SNe are in the range
$0.8-1.6$ (Maguire et al. 2010), hence this supports a host galaxy extinction of
around $A_{V}\simeq 4-5$. The NIR colours  $J-K=0.9\pm0.2$ and $H-K=0.5\pm0.2$
also support a large extinction for SN~2004am. Given the evidence that
low-luminosity SNe show a large NIR excess which changes rapidly at around
100 days (Pastorello et al. 2009), the $JHK$ colours are compatible with
extinctions in the range  $A_{V}\simeq3-6$. 

\subsection{Super-star cluster M~82-L and the progenitor of SN~2004am}
\label{ssc-progenitor}
The HST/ACS WFC F814W ($I$-band) image (0.05 arcsec pixel$^{-1}$) of M~82 obtained
on 2006 March 27 was aligned with the HST/ACS WFC F814W image from 2004 February 9.
We measured the centroid positions for 30 point-like sources visible in both
the images using the IRAF APPHOT package. A general geometric transformation function
was derived for the pairs of coordinates using the IRAF GEOMAP task. The astrometric
uncertainty was estimated from the RMS of the residuals from fitting the transformation
function to the data points. The uncertainties in both $x$ and $y$ estimated this way
were about 4 milliarcsec (mas). 

To measure the position of SN~2004am we first subtracted the aligned 2006 image from the
2004 image using the ISIS2.2 package. This revealed SN~2004am as a positive saturated
point-source at the location of M~82-L. To minimise the biasing effects of the saturation, the
position of the supernova was measured in the subtracted image making use of PSF fitting
within the SNOOPY\footnote{SNOOPY, originally presented in Patat (1996), has been im-
plemented in IRAF by E. Cappellaro. The package is based on DAOPHOT, but optimised for SN
magnitude measurements.} package based on IRAF's DAOPHOT. We also applied three different
methods within the IRAF's APPHOT package viz. centroiding, 2 dimensional Gaussian fitting,
and optimal filtering in order to estimate a conservative uncertainty for the position of SN~2004am.
The standard deviation of the four different measurements was found to be 20 mas in $x$ and
142 mas in $y$ (the large uncertainty in the $y$ direction was due to a strong spike in the
PSF). We, therefore, find that the supernova is located 32$\pm$20 mas in $x$ and 19$\pm$142
mas in $y$ from the centroid position of the SSC in the ACS image (see Fig. 1).

Higher resolution images of the field around M~82-L and M~82-F are available from the HST/ACS HRC
(0.025 arcsec pixel$^{-1}$) and we have also employed the F814W filter image in measuring the
clusters radial profile. These were taken on 2002 June 2007 and presented in \citeauthor{2007MNRAS.379.1333B} (2007).
In the high resolution ACS HRC F814W image the SSC is resolved, but it has a fairly sharp core compared
to the nearby M~82-F and two other nearby, fainter clusters (\citeauthor{2007MNRAS.379.1333B} 2007). The core has a half width
at half maximum (HWHM) of approximately 30 mas, but the wings of the radial profile are broad
and the half-width at zero intensity is 380 mas. M~82-F by comparison has a broader core with a
HWHM of 70 arcsec, but a similar spatial extent of approximately 400 mas. The best measured
position for SN 2004am therefore falls within the half-light radius of the cluster M~82-L and
we conclude that SN~2004am is spatially coincident with the SSC M~82-L and that its progenitor
was most likely a cluster member.

\begin{figure*}
\begin{center}
\includegraphics[width=175mm, angle=0, clip] {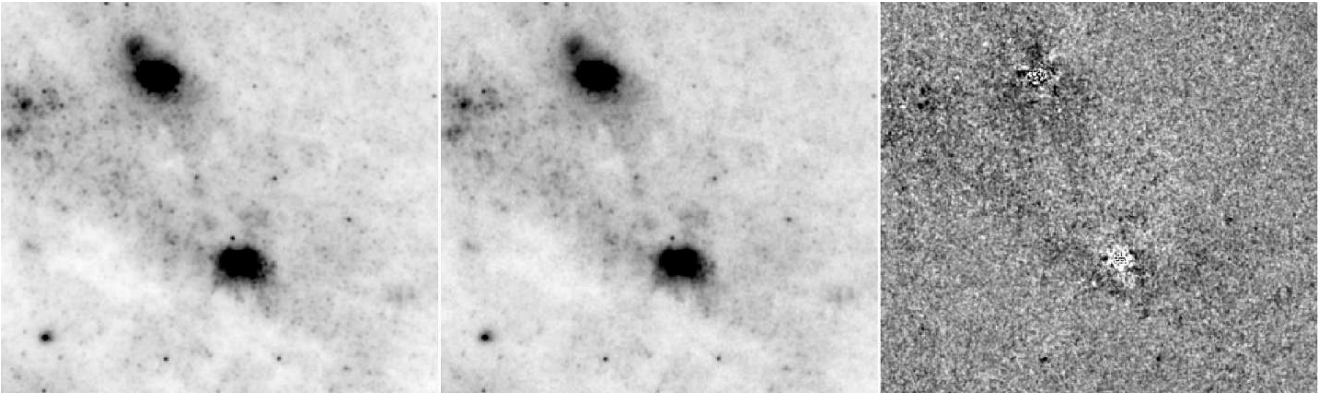}
\caption{Sections of ACS/HRC F814W images (North is Up and East to the Left)
showing the SSCs M~82-L (to the south-west from the centre)
and M~82-F obtained on 5 July 2004 ({\bf left}) and 7 June 2002 ({\bf middle})
and the subtraction between the two obtained using ISIS2.2 ({\bf right}). The
first two are shown with inverted intensity scale. Positive image subtraction
residuals are visible in the subtracted image at the location of M~82-L but
no evidence of a point source indicating that the SN has faded below the
HST detection limit.}
\end{center}
\end{figure*}

M~82-L has been extensively studied in the optical and the NIR, with
the first attempt at an age determination offered by \citeauthor{1999MNRAS.304..540G} (1999). 
Their spectra of M~82-L only included red wavelengths, resulting in
a somewhat uncertain estimate of cluster age. However, they suggest
an age of around 60 Myrs, similar to that obtained for the M~82-F, as
these clusters display comparable Ca\,{\sc ii} triplet line strengths.
{\em Space Telescope Imaging Spectrograph} (STIS) spectra of M~82-L 
were analysed by \citeauthor{2006MNRAS.370..513S} (2006) with the spectral synthesis code
$starburst99$ (\citeauthor{1999ApJS..123....3L} 1999).
A spectroscopic age for the cluster of $65^{+70}_{-35}$\,Myrs and extinction of
$E(B-V)=1.87\pm0.39$ was derived. The large error bars arise mostly from the
fact the the $U$-band was not covered and hence extinction and age become
difficult to disentangle. 

The more recent study of \citeauthor{2008A&A...486..165L} (2008) presented a medium resolution ($R\simeq750$) and 
high signal-to-noise $0.8-2.4 \mu$m spectrum of M~82-L. They used the stellar
population synthesis code ({\sc P\'{E}GASE.2}), with a library of observed
red supergiant spectra to produce single stellar population spectra with
ages between  8 to 60 Myr. In all cases near-solar metallicity is assumed,
along with a Salpeter initial mass function (IMF). An impressive fit to the
$JHK$ spectra of M~82-L is obtained for a single stellar population of age $18^{+17}_{-8}$\,Myr.
The age of the cluster provides a strong constraint on the mass of the
progenitor star, if we assume that the cluster formed coevally and that the
cluster age is representative of that of the progenitor. 
\citeauthor{2009MNRAS.395.1409S} (2009) employed the stellar evolutionary tracks of 
\citeauthor{2004MNRAS.353...87E} (2004) to determine stellar masses of II-P progenitors. 
Using these tracks (to allow a consistent comparison) we find that an age of 
$18^{+17}_{-8}$\,Myr corresponds to lifetimes of stars of initial masses
$12^{+7}_{-3}$\,\msol. The models in the single stellar population analysis of \citeauthor{2008A&A...486..165L}
(2008) were those of \citeauthor{1993A&AS..100..647B} (1993) and the age-mass relationships
in both codes are quite similar. 

\begin{figure*}
\begin{center}
\includegraphics[width=85mm]{mattila_fig3a.eps}
\includegraphics[width=75mm]{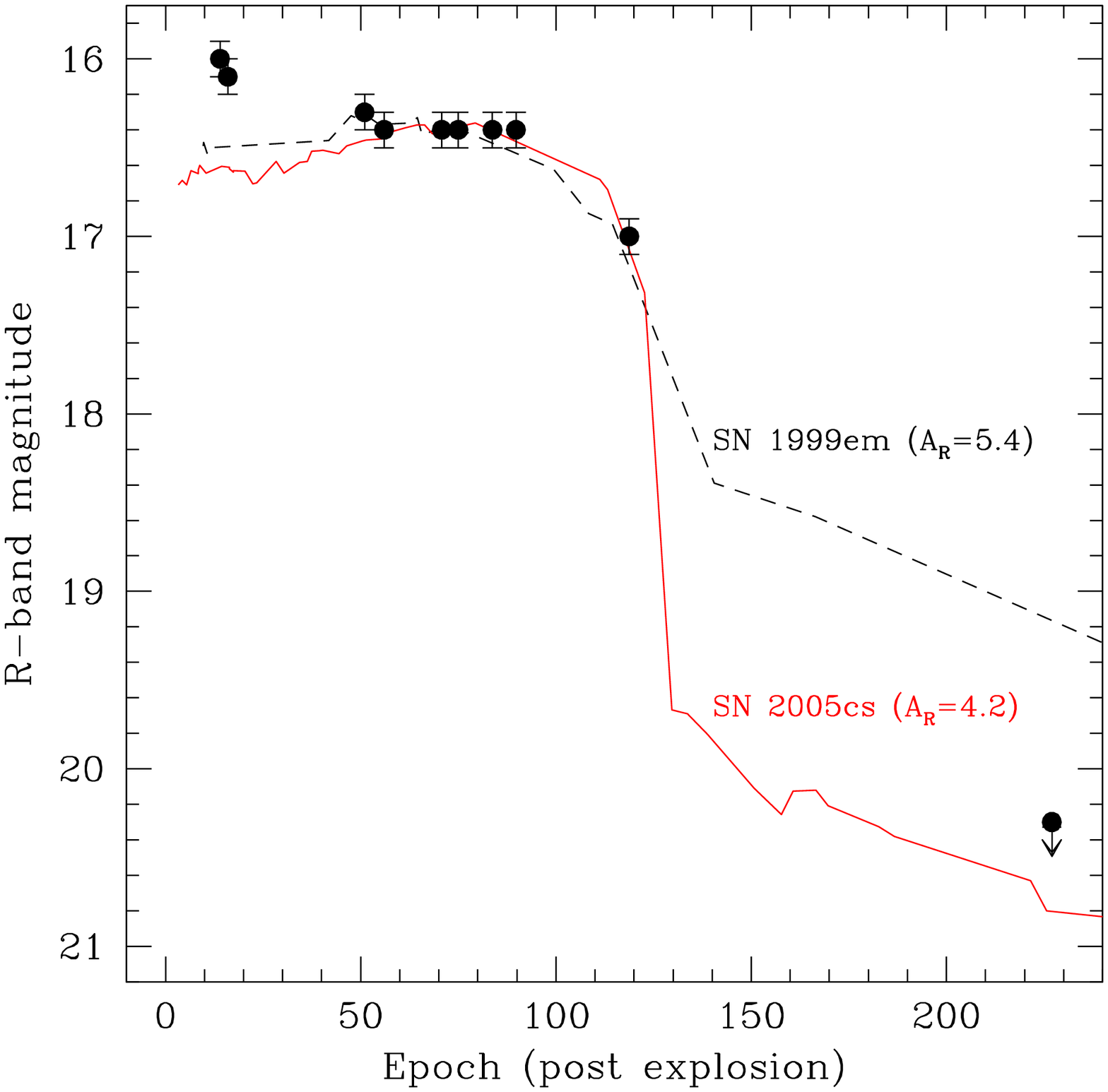}
\vspace{-0.3cm}
\caption{{\bf Left-Top :} 
A comparison of the SN~2004am $zJ$ spectra obtained on 2004 Mar 6
(black) with the spectrum of SSC M~82-L from Lan{\c c}on et al. (2008) shown in red (along with 
a Gaussian smoothed spectrum overlaid in green). The pre-discovery 
spectrum of M~82-L was scaled and subtracted from the 2004 Mar 6 spectrum
to leave a clean spectrum of SN~2004am. The major features are marked
at the positions of the probable P-Cygni absorption troughs. 
{\bf Left-Lower :}
A comparison of the SN~2004am $zJ$ spectrum with that of three other
well studied II-P SNe near the end of the plateau phase. All spectra
were dereddened and scaled to match the flux of SN~2004am at 1.3$\mu$m
(see text). As the spectra were taken with a diversity of resolutions, all were 
broadened (with a Gaussian FWHM of 40\AA) and rebinned (to 20\AA pix$^{-1}$)
to the lowest resolution for a meaningful comparison. 
The SN~2004am spectrum was dereddened with $A_{V} = 3.7$ and $R = 2.4$
as found for M~82-L by Lan{\c c}on et al. (2008). 
{\bf Right :}
The light curve of SN~2004am compared with the $R$-band light curves 
of two type II-P SNe. The comparison light curves were scaled to a distance of
3.3\,Mpc and dimmed to correspond to host galaxy extinctions of A$_{\rm R}$
= 5.4 and 4.2 for SN 1999em and 2005cs, respectively. The plateau points are unfiltered
CCD magnitudes from Singer, Pugh \& Li (2004), and the upper limit at
241 days is an estimated limit from the F814W HST image. An epoch of explosion of 14 days
before the first observations by LOSS was adopted for SN 2004am to yield a good match with the
post-plateau drop in the light curves of SNe 1999em and 2005cs.} 
\end{center}
\end{figure*}

\section{SN~2008iz and the 43.78+59.3 transient - Observations and results}
\subsection{Observations and data reduction}
The sites of SN~2008iz and the 43.78+59.3 transient were observed on 2009 Jun 11 
using the Near-Infrared Imager (NIRI) with the ALTAIR AO system
on the Gemini-North Telescope, as part of our SN progenitor identification 
program\footnote{GN-2009A-Q-32}. The f/14 camera was used, giving 0.05\arcsec 
pixels over a 51\arcsec $\times$ 51\arcsec field of view. While typically in 
AO observations of SNe, the SN itself is used as a natural guide star (or a 
tip-tilt star together with a laser guide star), this is only possible for 
bright ($V < 15$ mag) SNe. Hence for these observations obtained in the
natural guide star mode we used the nearby cluster M~82-F as the natural guide star.
During the observations the seeing was exceptionally good, mostly FWHM $<$ 0.4''
(as measured by ALTAIR), yielding a reasonable AO correction quality (FWHM $\sim$ 0.2''),
despite M~82-F being an extended source. An integration time of 60 sec was used
per pointing and separate sky frames were obtained. The image shown in Fig. 4 has
a total on-source integration time of 30 minututes and was reduced using standard
techniques (sky subtraction and masking of bad pixels) for NIR imaging with the
{\sc iraf.gemini} package.

As neither SN~2008iz nor the 43.78+59.3 transient were immediately visible at their 
positions from the radio observations (\citeauthor{Fra09} 2009), we attempted to perform 
image subtraction to identify any faint transient source or low level variability at 
the site of either transient. We obtained pre-explosion observations from 
HST+NICMOS via MAST\footnote{http://archive.stsci.edu} to use as reference images
for the image subtraction. For SN~2008iz, the images used were a 448s exposure 
with the NIC2 camera (pixel scale 0.075\arcsec/pix) in the F222M filter and a
1408s exposure in the F237M filter obtained using the same camera. For the 
43.78+59.3 transient, the templates used were a 448s exposure in F222M and a 
1408s exposure in F237M. All the NICMOS images were taken on 1998 April 12; 
the pipeline-reduced products from the HST archive were used.

\subsection{Alignment and SN identification}
As the field of view of both the NIRI (51\arcsec $\times$ 51\arcsec) and NICMOS 
(19\arcsec $\times$ 19\arcsec) images are small compared to 
the angular size of M~82, we employed a bootstrap technique to obtain accurate 
astrometry for our data. We first identified SDSS sources in an $F814W$-filter 
$HST$ ACS mosaic (Mutchler et al. 2007) of M~82, 
which allowed us to calibrate the world coordinate system (WCS) of this image 
to a high degree of accuracy, and subsequently used this frame as an astrometric 
reference for all other images.

The ACS mosaic of M~82 was downloaded from the $HST$ MAST high level science 
products webpages\footnote{http://archive.stsci.edu/prepds/m82/}. This mosaic
image has been corrected for the geometric distortion of ACS, and its internal
astrometric accuracy is extremely good, with residuals in the alignment of 
the input tiles to the mosaic of $\sim$0.1 pixels, corresponding to 5 mas. We 
identified 51 point sources which are in the Sloan catalog 
(\citeauthor{2009ApJS..182..543A} 2009) with an $r$ magnitude $<$ 21 mag. The 
coordinates of these sources were measured with the {\sc iraf phot} task. 
Using the pixel coordinates of the sources, together with the Sloan catalog 
celestial coordinates, we refit the WCS for the ACS mosaic using {\sc iraf 
ccmap} and {\sc ccsetwcs}. A general transformation was allowed, consisting of 
translation, rotation, scaling and a polynomial term. Three sources were 
discarded at this stage as outliers. The rms error in fitting the new WCS was 
48 and 41 mas in R.A. and Dec., respectively.

Next, we identified 26 sources in common between our NIRI $K$-band image and 
the ACS $F814W$-filter mosaic. We measured the pixel coordinates of these as 
before with {\sc iraf phot}, and fitted the matched coordinate list with a 
general transformation using {\sc iraf geomap}. We rejected two sources as clear 
outliers from the fit. The rms error in the fit was 20 and 22 mas in R.A. and 
Dec., respectively. The NIRI image was transformed to match the ACS image, and 
the two frames compared by eye to verify the transformation.

\begin{figure*}
\begin{center}
\includegraphics[width=150mm]{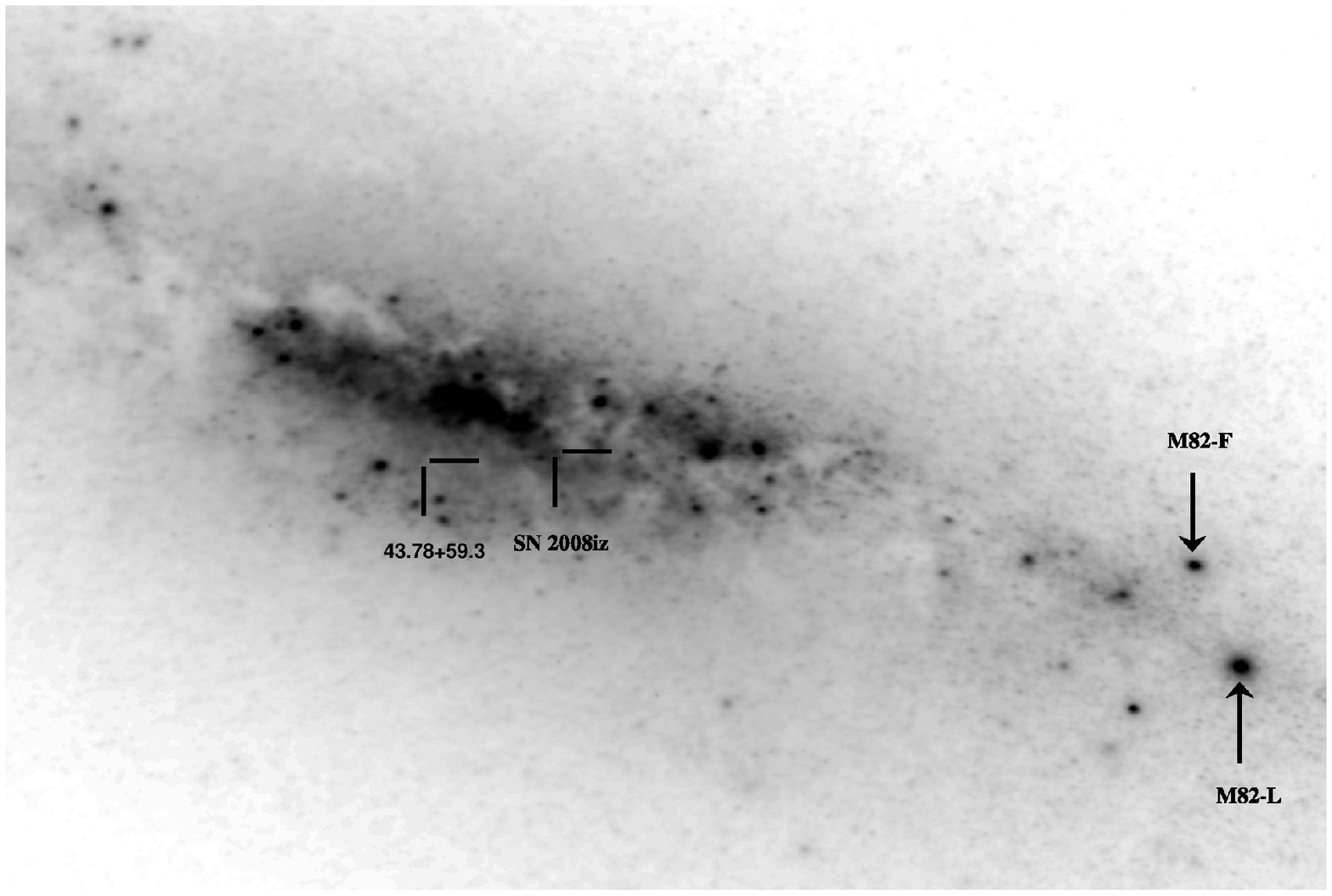}
\vspace{-0.3cm}
\caption{A 50''$\times$33'' region of the K-band Gemini-N Altair/NIRI image of the M~82 nuclear regions (North is up and East to the left).
The locations of SN~2008iz, the 43.78+59.3 transients and the SSCs M~82-L and M~82-F are indicated.} 
\end{center}
\end{figure*}

Using the Sloan-calibrated WCS, and using the values of R.A. and Dec. from radio
observations given in 
Table \ref{t_coords} we identified the pixel coordinates of SN~2008iz and the 
43.78+59.3 transient in the ACS image.
These coordinates were then transformed to the NIRI image using the 
transformation determined previously with the {\sc geoxytran} task.

Unfortunately, the site of SN~2008iz and the 43.78+59.3 transient are on different 
NICMOS images, and so each had to be aligned to the NIRI image separately. In 
each case, we matched between ten and twenty common sources, and transformed 
the NICMOS image to the pixel coordinates of the NIRI image using {\sc iraf 
geomap} and {\sc geotran}.

For SN~2008iz, we used the {\sc isis}2.2 package
to convolve the $F222M$- and $F237M$-filter NICMOS images to match the PSF and 
flux level of the NIRI $K$-band image. We then subtracted the convolved 
$F222M$ and $F237M$ images from the $K$ image, and searched for residual flux 
at the site of the radio SN. As an additional test we subtracted the 
$F222M$ image from the $F237M$ image to see if a colour difference could 
produce a source.

The results of these subtractions are shown in Figures 5 and 6. As can be seen,
there is a clear detection of a source co-incident with the radio coordinates of
SN~2008iz in the 
subtractions between the NIRI and NICMOS images. The source in the difference 
image consists of positive flux, implying a brightening since April 1998. The 
source is clearest in the subtraction between NIRI $K$ and NICMOS $F222M$ 
images (which is unsurprising given the good match between the bandpasses of 
the filters, as shown in Figure \ref{fig_filter}), although it is still 
significant in the $F237$ filter. No source is detected at the site of the SN 
in the subtraction between the $F222M$ and $F237M$ filters, indicating that
there is not a point source with a large $F222M-F237M$ colour at this location 
that could account for the source in the other subtractions.

We repeated the same procedure for the 43.78+59.3 transient, again finding a 
source at the radio position of the transient in the difference image. As the 
detection was not as clear as in the case of SN~2008iz, we repeated the 
convolution and subtraction process using {\sc hotpants}\footnote{http://www.astro.washington.edu/users/becker/hotpants.html}, 
but obtained the same result.

We measured the pixel coordinates of both sources in the difference images 
obtained with ISIS and the F222M template image using the optimal filter centering 
algorithm in {\sc iraf phot}. In the case of SN~2008iz, we find an offset of 
0.68 pixels (34 mas) between the expected and measured position of the 
transient, while for the 43.78+59.3 transient there is a 1.62 pixel (81 mas) 
offset. While this is slightly outside the combined rms error of the alignment 
(70 mas), we nonetheless consider the association convincing given that the 
transient source is detected at a relatively low signal to noise. We identified nine other
point sources which appeared in a 12\arcsec $\times$12 \arcsec region in the 
difference image for the 43.78+59.3 transient; from this we calculate a probability of
a variable or transient source coincident to less than 81 mas as $\sim$0.1 per cent. 

\begin{figure*}
\subfigure[Gemini+NIRI ($K$) post-explosion]{
\includegraphics[width=44.5mm,angle=0]{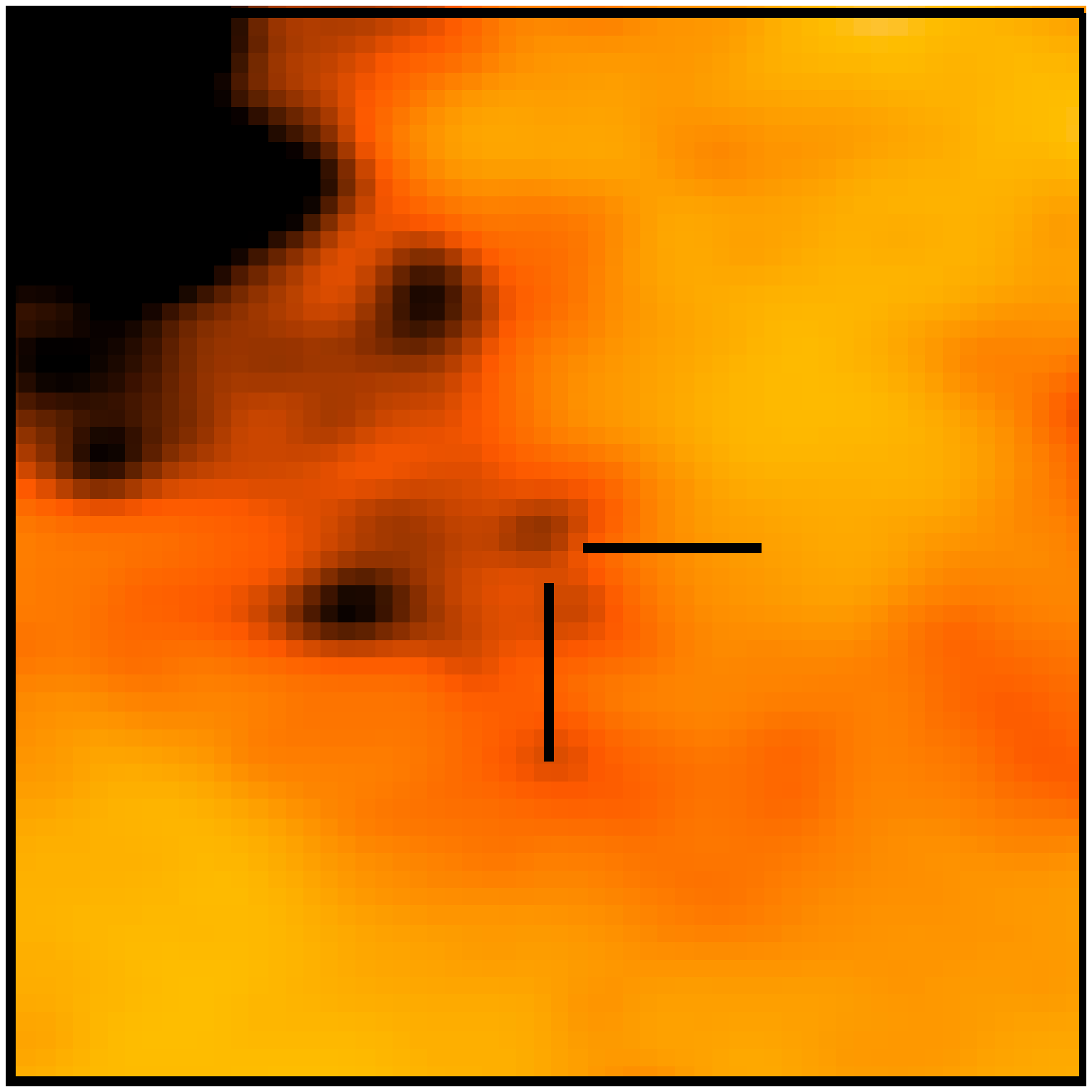}
\label{fig2:subfig1}
}
\subfigure[HST+NICMOS ($F222M$) pre-explosion, transformed to match NIRI image]{
\includegraphics[width=44.5mm,angle=0]{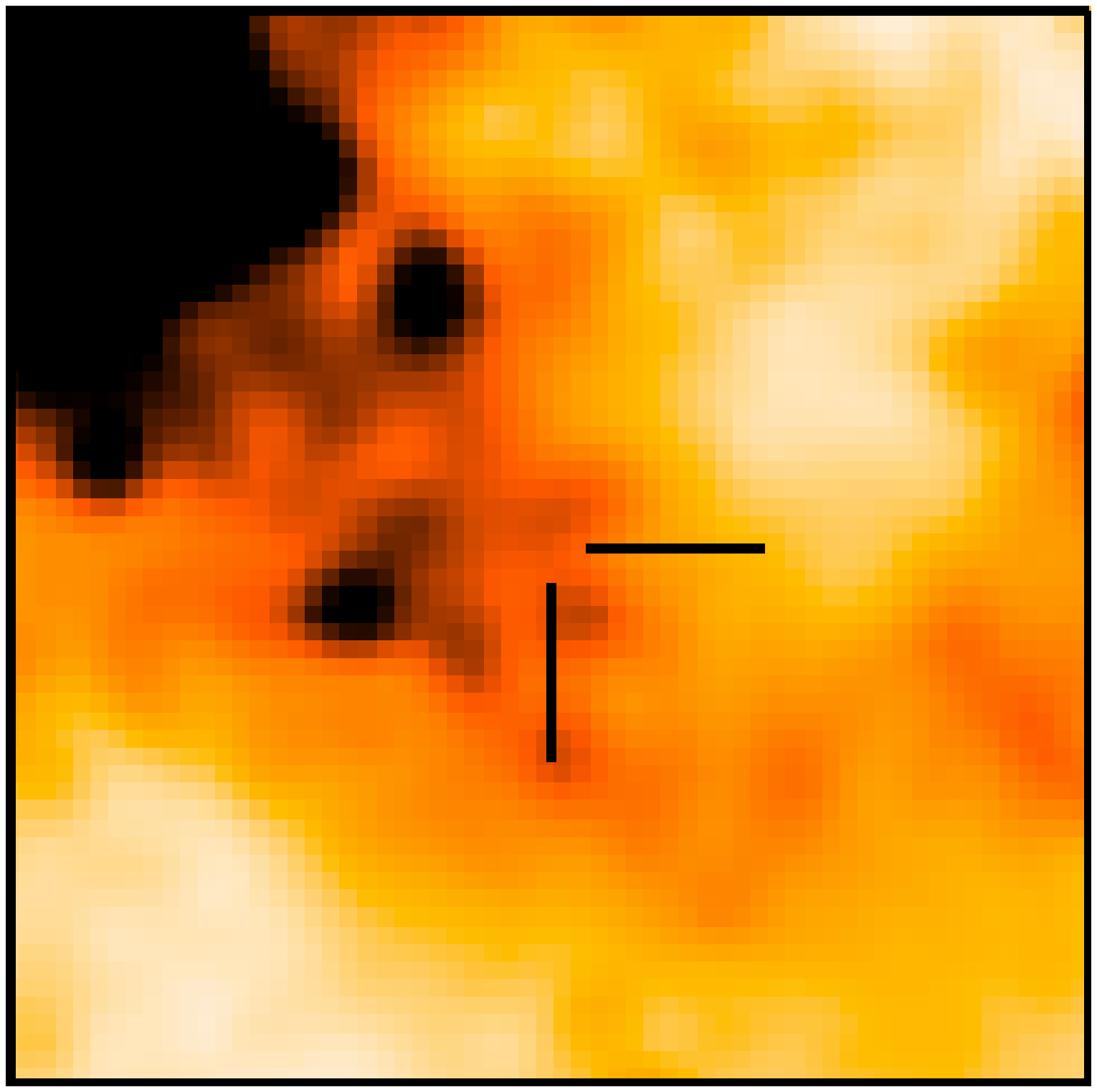}
\label{fig2:subfig2}
}
\subfigure[Subtraction of \ref{fig2:subfig2} from \ref{fig2:subfig1} using {\sc isis}]{
\includegraphics[width=44.5mm,angle=0]{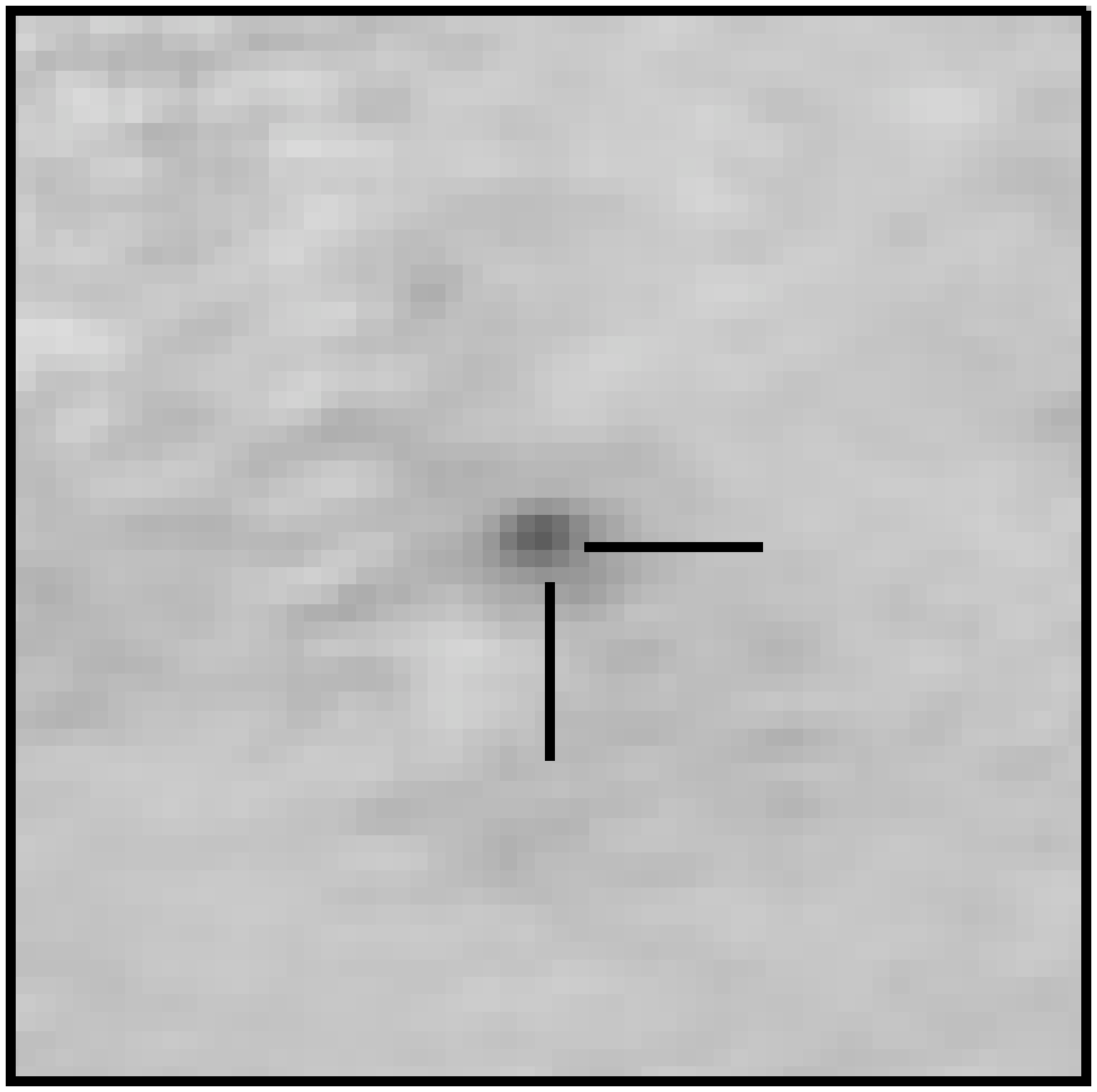}
\label{fig2:subfig3}
}
\subfigure[Gemini+NIRI ($K$) post-explosion]{
\includegraphics[width=44.5mm,angle=0]{mattila_fig5a.eps}
\label{fig2:subfig4}
}
\subfigure[HST+NICMOS ($F237M$) pre-explosion, transformed to match NIRI image]{
\includegraphics[width=44.5mm,angle=0]{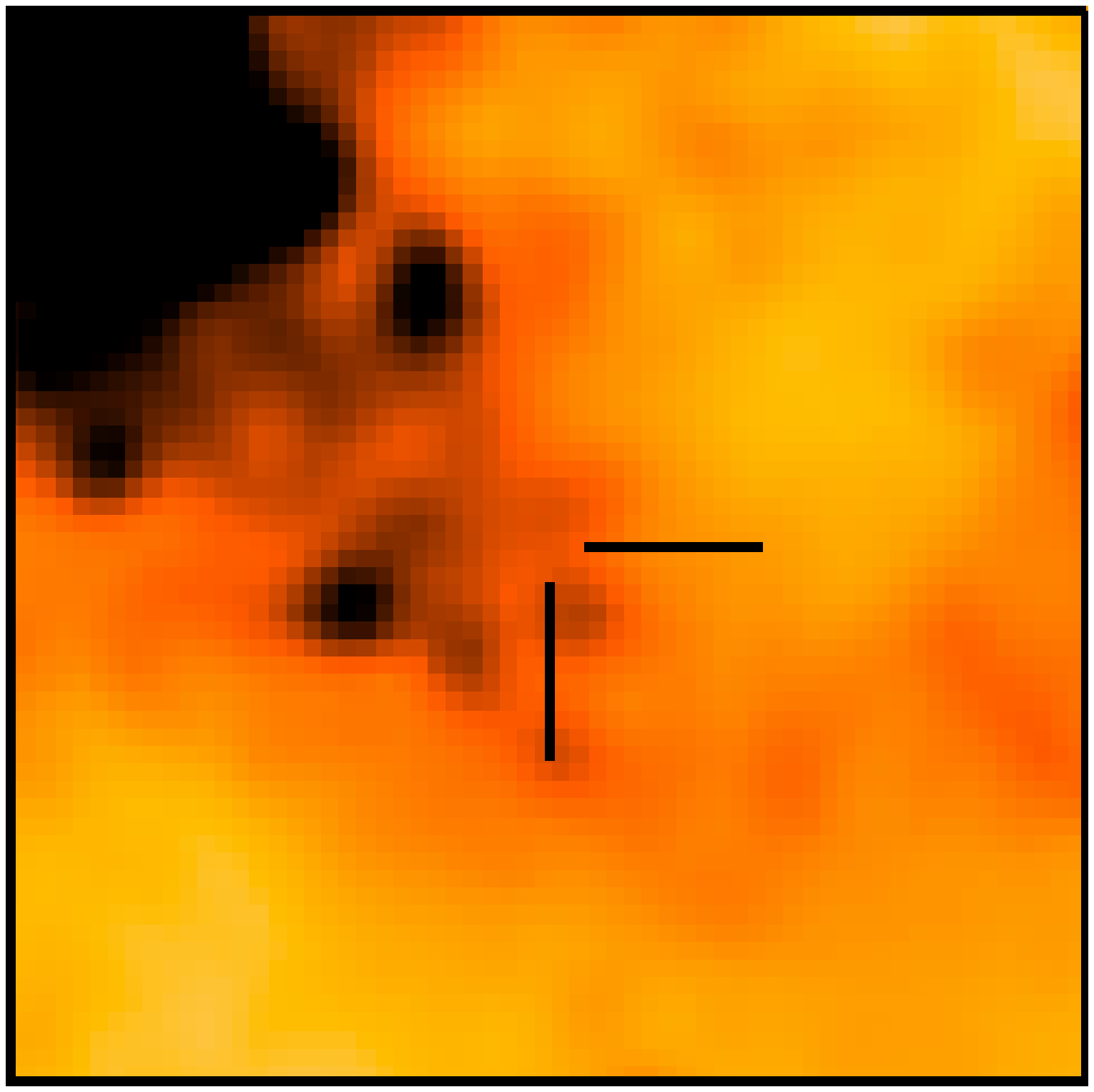}
\label{fig2:subfig5}
}
\subfigure[Subtraction of \ref{fig2:subfig5} from \ref{fig2:subfig4} using {\sc isis}]{
\includegraphics[width=44.5mm,angle=0]{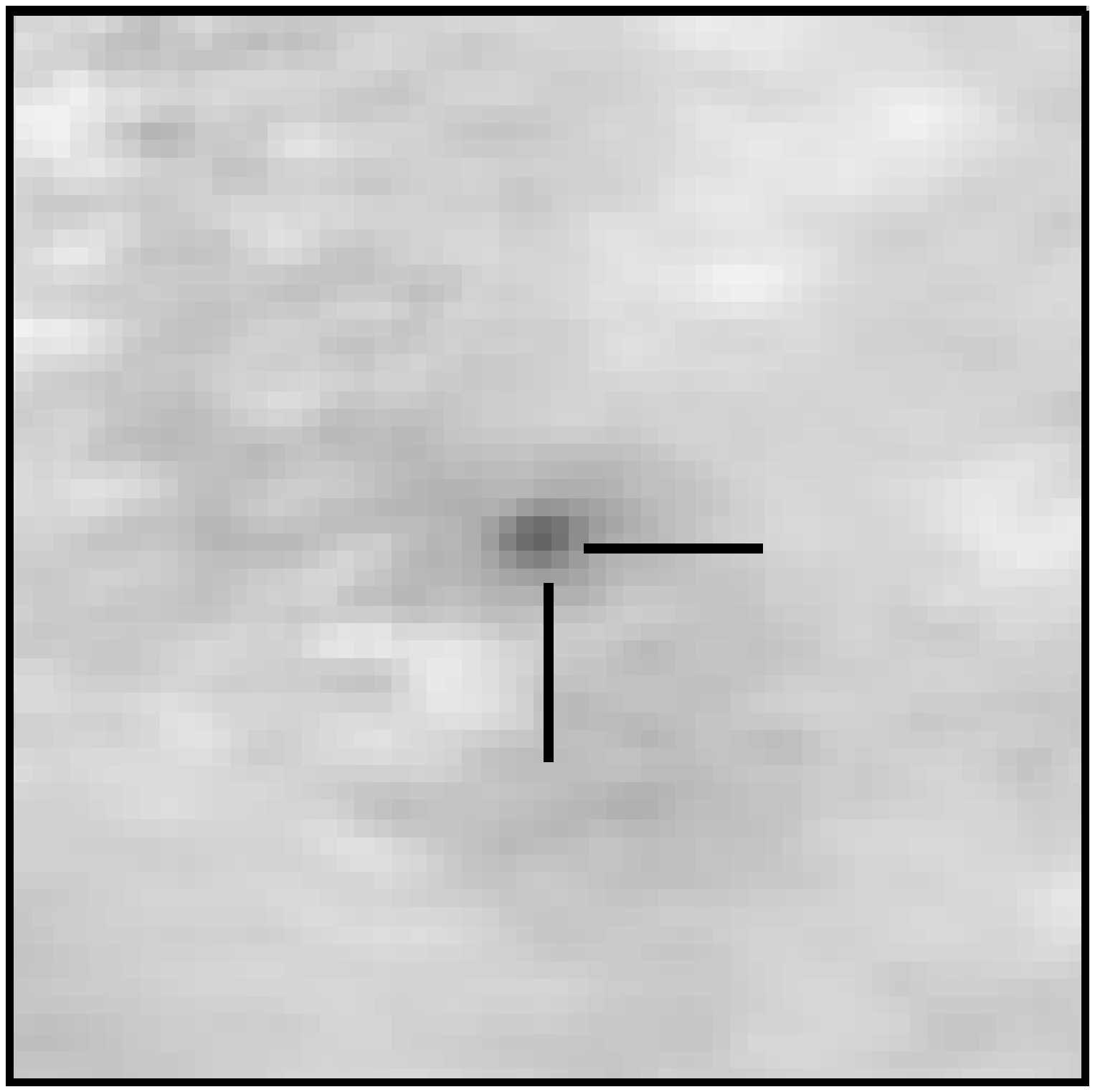}
\label{fig2:subfig6}
}
\subfigure[HST+NICMOS ($F222M$) pre-explosion]{
\includegraphics[width=44.5mm,angle=0]{mattila_fig5b.eps}
\label{fig2:subfig7}
}
\subfigure[HST+NICMOS ($F237M$) pre-explosion]{
\includegraphics[width=44.5mm,angle=0]{mattila_fig5e.eps}
\label{fig2:subfig8}
}
\subfigure[Subtraction of \ref{fig2:subfig7} from \ref{fig2:subfig8} using {\sc isis}]{
\includegraphics[width=44.5mm,angle=0]{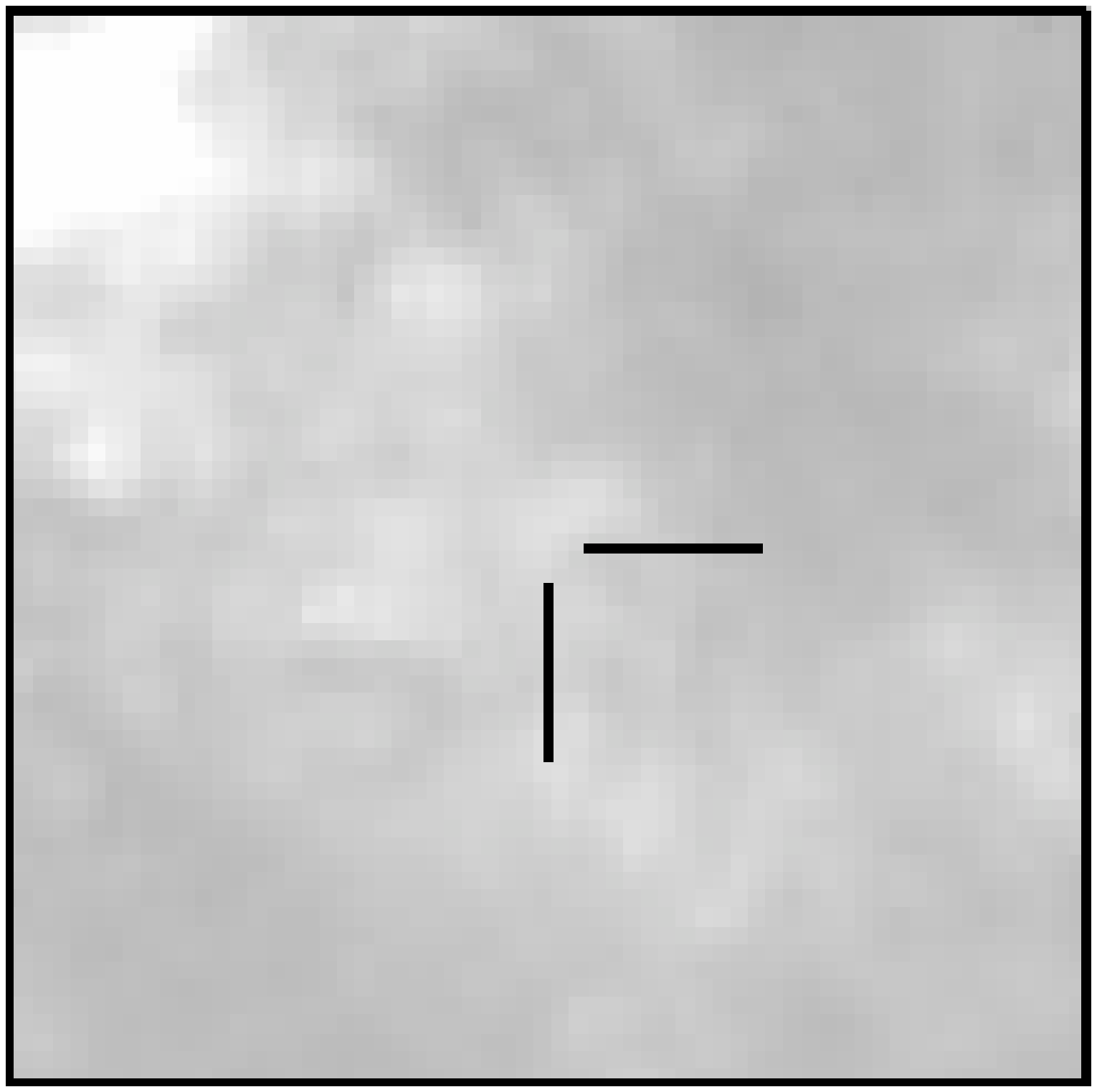}
\label{fig2:subfig9}
}
\caption{Gemini post- and HST pre-explosion images of the site of SN~2008iz, 
together with difference images obtained with {\sc isis}. In all cases the 
images are centred on the radio coordinates of the transient, with North up and East left. 
All images are 3\arcsec $\times$ 3\arcsec.}
\end {figure*}

\begin{figure*}
\subfigure[Gemini+NIRI (Ks) post-explosion]{
\includegraphics[width=41.4mm,angle=270]{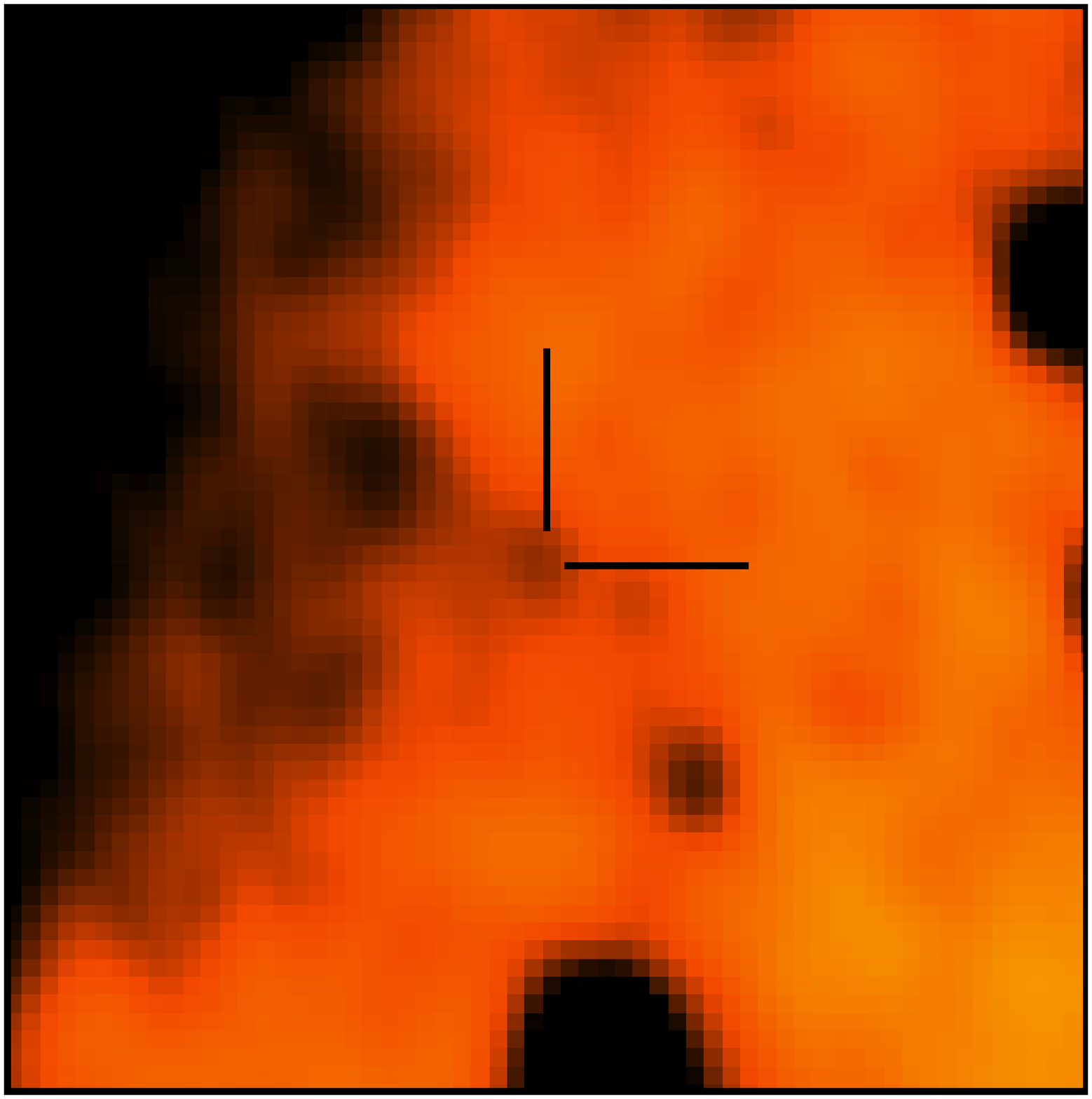}
\label{fig:subfig1}
}
\subfigure[HST+NICMOS (F222M) pre-explosion, transformed to match NIRI image]{
\includegraphics[width=41.4mm,angle=270]{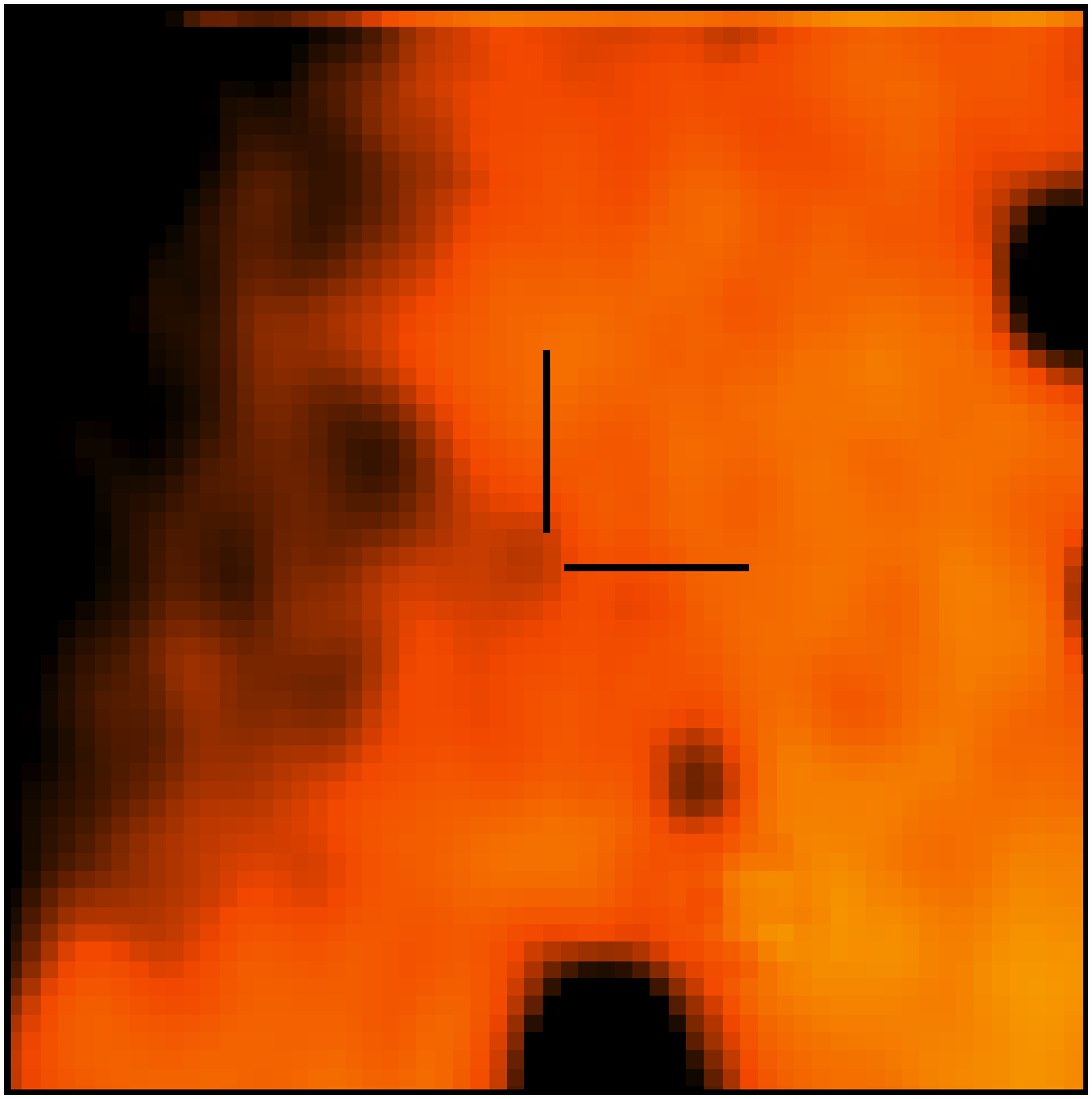}
\label{fig:subfig2}
}
\subfigure[Subtraction of \ref{fig:subfig2} from \ref{fig:subfig1} using ISIS]{
\includegraphics[width=41.4mm,angle=270]{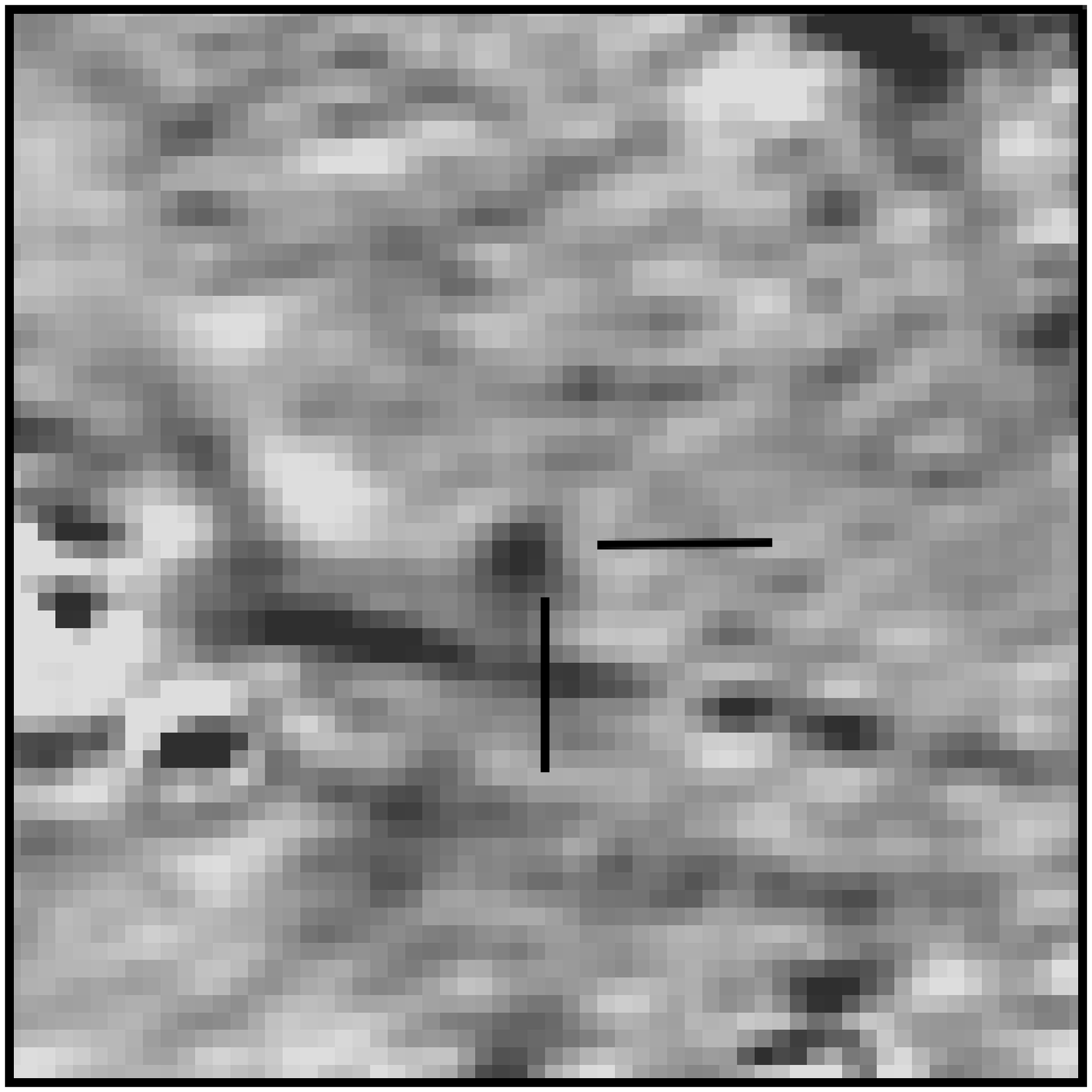}
\label{fig:subfig3}
}
\subfigure[Subtraction of \ref{fig:subfig2} from \ref{fig:subfig1} using {\sc hotpants}]{
\includegraphics[width=41.4mm,angle=270]{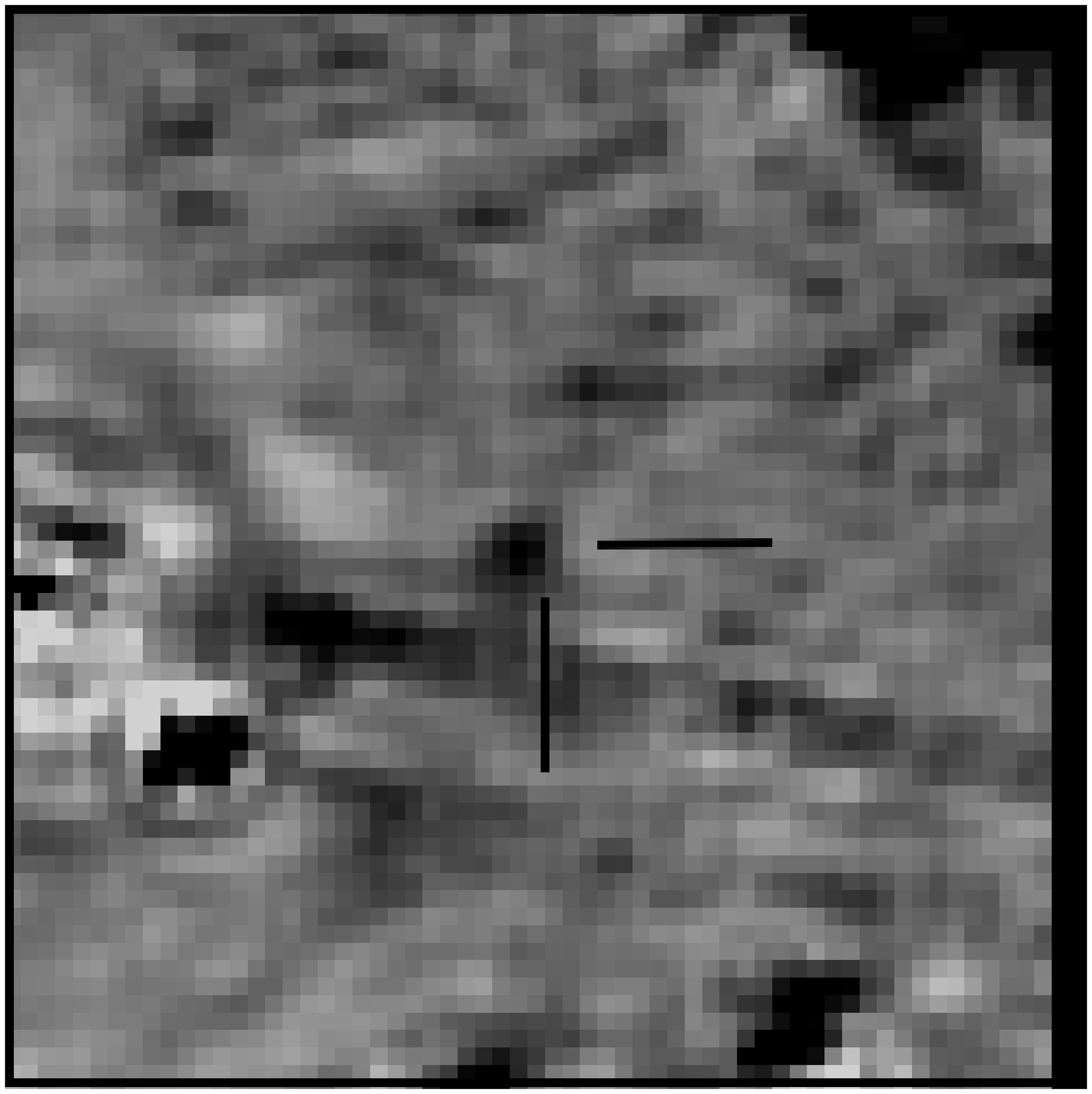}
\label{fig:subfig4}
}
\subfigure[Gemini+NIRI (Ks) post-explosion]{
\includegraphics[width=41.4mm,angle=270]{mattila_fig6a.eps}
\label{fig:subfig5}
}
\subfigure[HST+NICMOS (F237M) pre-explosion, transformed to match NIRI image]{
\includegraphics[width=41.4mm,angle=270]{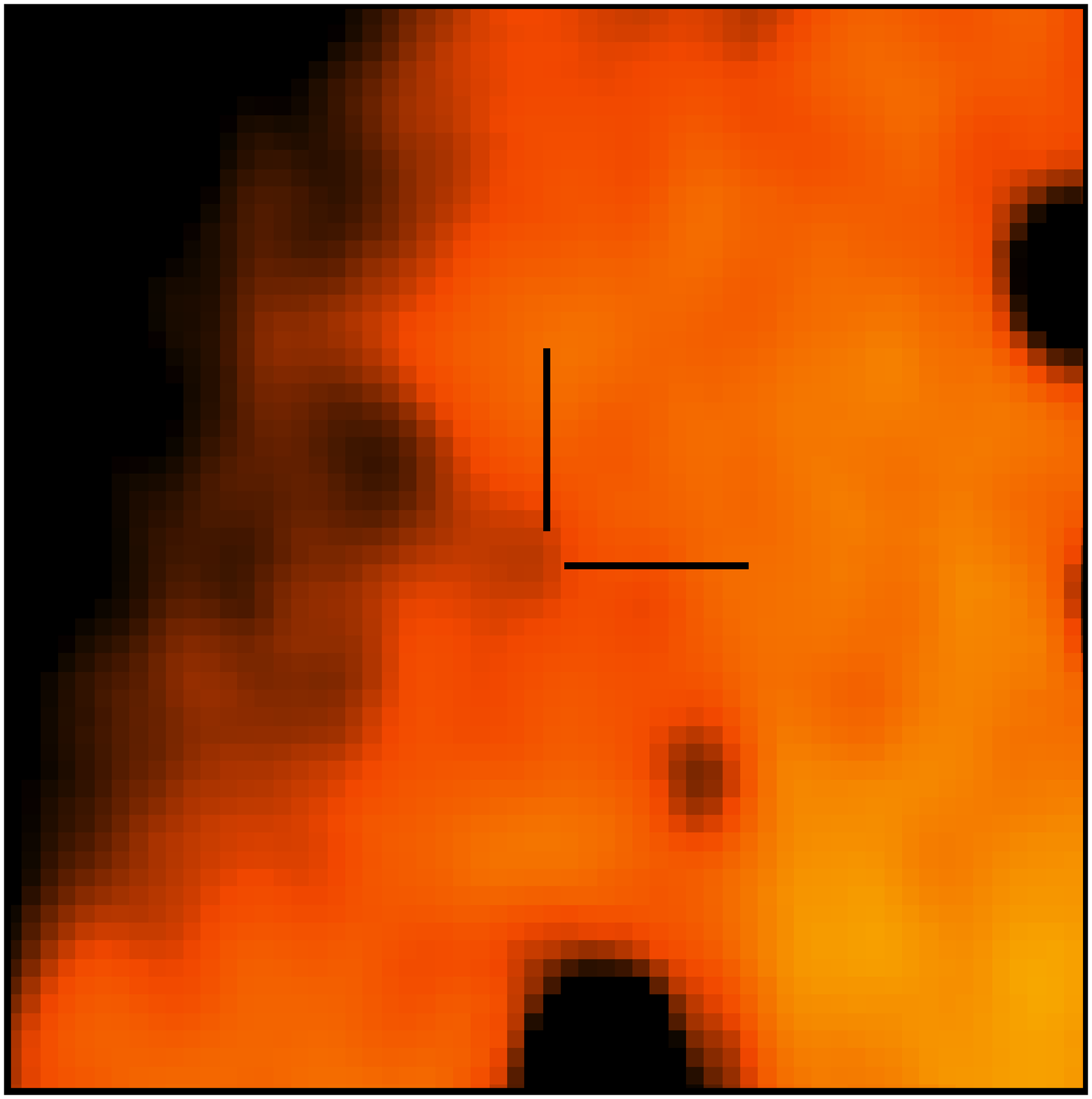}
\label{fig:subfig6}
}
\subfigure[Subtraction of \ref{fig:subfig6} from \ref{fig:subfig5} using ISIS]{
\includegraphics[width=41.4mm,angle=270]{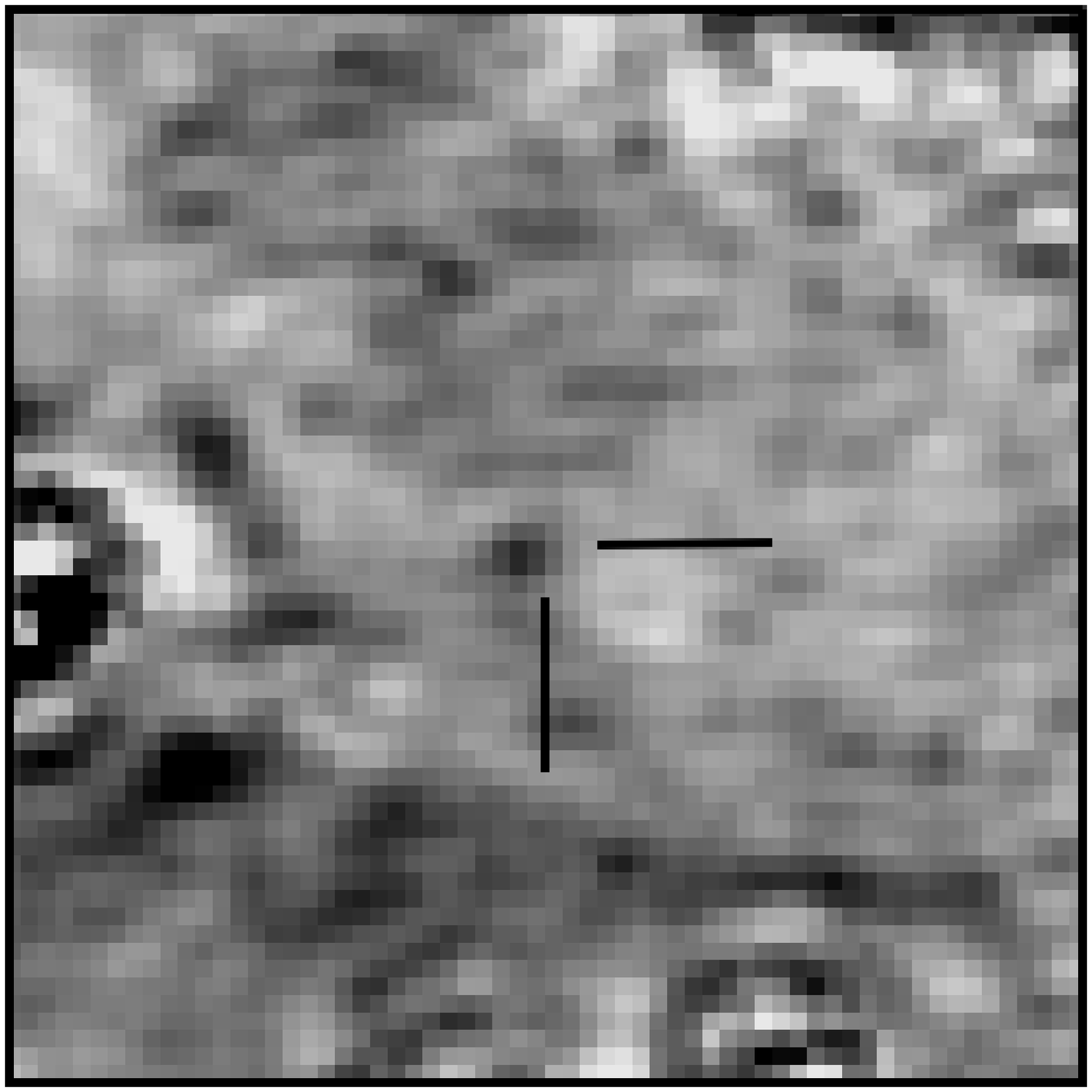}
\label{fig:subfig7}
}
\subfigure[Subtraction of \ref{fig:subfig6} from \ref{fig:subfig5} using {\sc hotpants}]{
\includegraphics[width=41.4mm,angle=270]{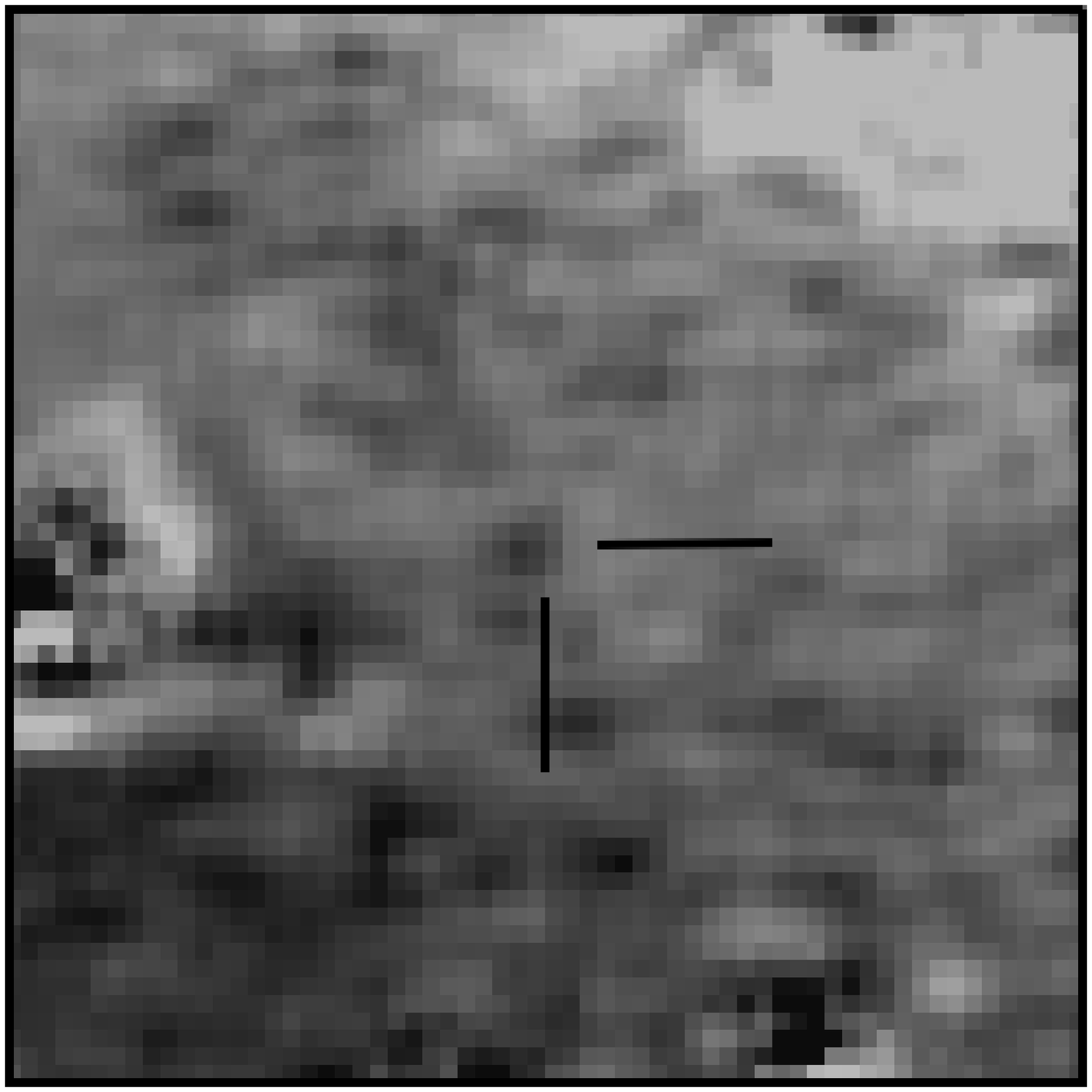}
\label{fig:subfig8}
}
\subfigure[HST+NICMOS (F237M) pre-explosion, transformed to match NIRI image]{
\includegraphics[width=41.4mm,angle=270]{mattila_fig6f.eps}
\label{fig:subfig9}
}
\subfigure[HST+NICMOS (F222M) pre-explosion, transformed to match NIRI image]{
\includegraphics[width=41.4mm,angle=270]{mattila_fig6b.eps}
\label{fig:subfig10}
}
\subfigure[Subtraction of \ref{fig:subfig10} from \ref{fig:subfig9} using ISIS]{
\includegraphics[width=41.4mm,angle=270]{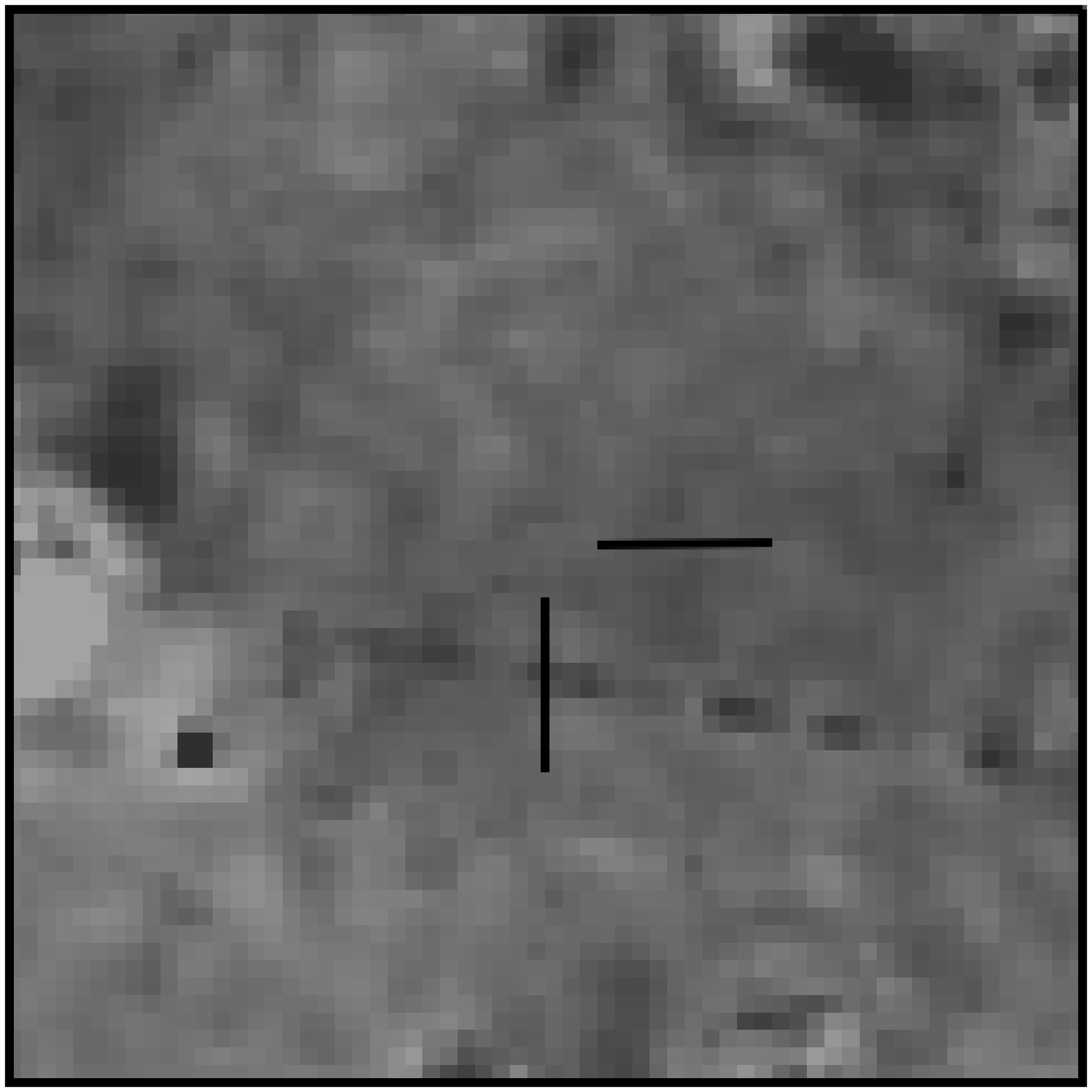}
\label{fig:subfig11}
}
\subfigure[Subtraction of \ref{fig:subfig10} from \ref{fig:subfig9} using {\sc hotpants}]{
\includegraphics[width=41.4mm,angle=270]{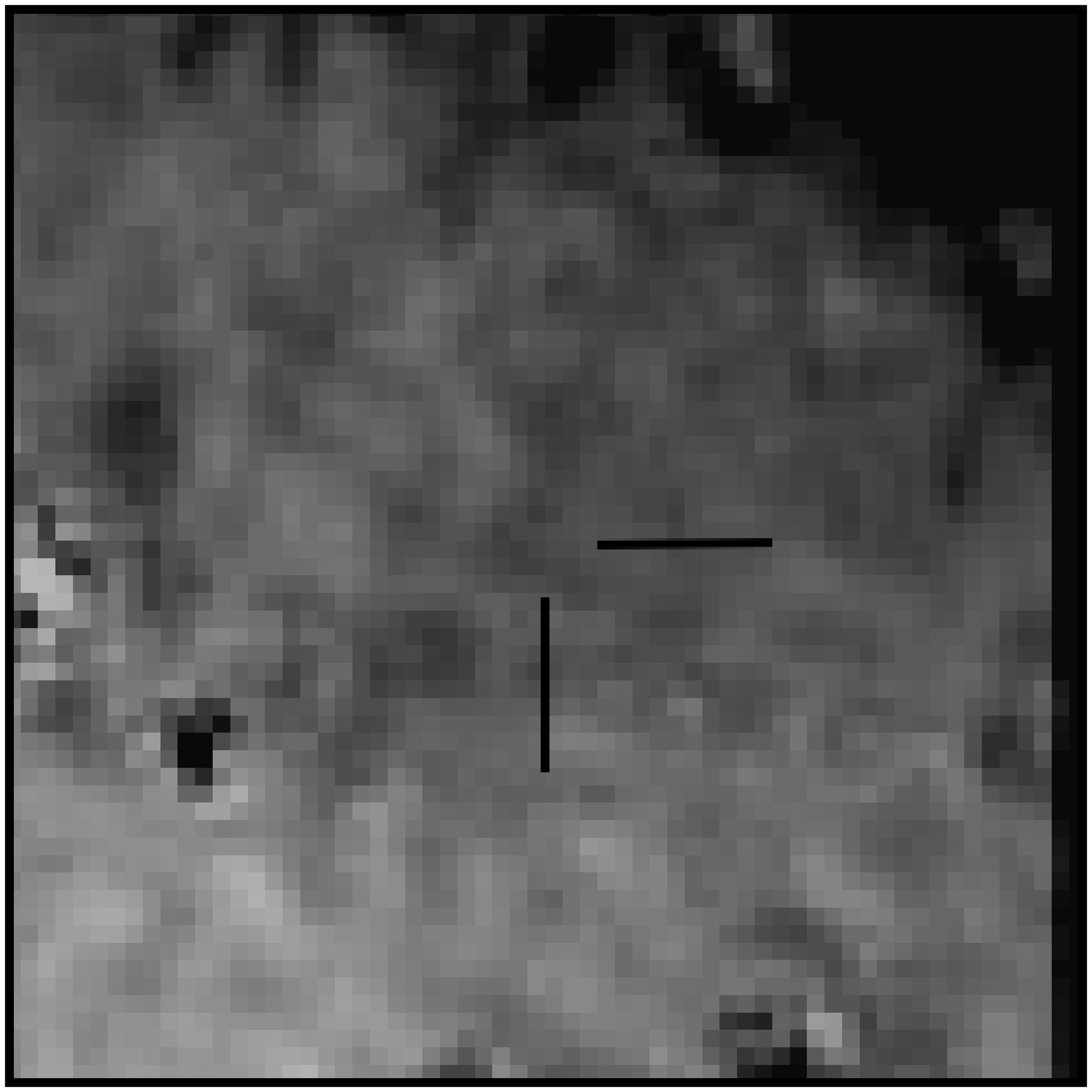}
\label{fig:subfig12}
}
\caption{Gemini post- and HST pre-explosion images of the site of the 43.78+59.3 transient, together with difference images obtained using {\sc isis} and {\sc hotpants}. In all cases the images are centred on the radio coordinates of the transient, with North up and East left. All images are 3\arcsec $\times$ 3\arcsec.}
\end {figure*}

\subsection{Photometry of SN~2008iz and the 43.78+59.3 transient}

For SN~2008iz and the 43.78+59.3 transient, we performed both PSF-fitting and
aperture photometry in the $K$-$F222M$ difference images. To set the zero point for
the photometry, we measured the most isolated and point-like SSCs
6, c, d, F and L in the NIRI frame, adopting their $F222M$ magnitudes from \citeauthor{2003ApJ...596..240M} (2003).
As no bright enough individual stars were covered within the NIRI field of view, also the PSF was
determined using the SSCs. According to \citeauthor{2003ApJ...596..240M} (2003) these SSCs have
their projected half-light radii less than $\sim$0.09'', i.e., are unresolved in our NIRI images
with spatial resolution of FWHM $\sim$0.2'' and are within $\sim$30'' from SN~2008iz and the
43.78+59.3 transient. The average value and the standard deviation of the zeropoints obtained with the
different clusters were adopted to be used for the photometric calibration and its associated
uncertainty. We also attempted to use 2MASS sources in the NIRI field of view to set the zeropoint,
but unfortunately there was an insufficient number of isolated sources with reliable 2MASS $K$-band
magnitudes covered by the field of view. As a consistency check, photometry was also obtained from a subtracted
image where the NIRI frame was scaled to match the flux units of the NICMOS template prior to the
subtraction. Using the PHOTFLAM value in the FITS header of the latter yielded a $K$ magnitude for
SN~2008iz within $\sim$0.1 mag from the one obtained using the SSCs.

Using PSF fitting for SN~2008iz gives a magnitude of $F222M$ ($\sim$$K$) = 15.91$\pm$0.16
where the error is dominated by the uncertainty in the zeropoint magnitude ($\pm$0.14 mag).
The photometric uncertainty was estimated via PSF-fitting to artificial sources placed close
to the SN position after subtracting the PSF-fit at the SN position. For comparison,
using a photometric aperture with an 8 pixel radius for SN~2008iz gives a magnitude of
16.03. In this case the zeropoint was obtained using aperture photometry (8 pixel radius)
pf the five SSCs. Varying the photometric aperture by a few pixels causes the photometry of SN~2008iz
to vary by up to 0.2 mag. The measured magnitude is also dependent on the region used to
measure the sky background at the SN location. The convolution used in the image subtraction
will also affect the noise properties of the subtracted image. For these reasons, we adopt a 
conservative error of $\pm$0.3 mag for our aperture photometry which is larger than would be
implied by Poissonian statistics.

For the 43.78+59.3 transient we obtain a magnitude of $F222M$ = 17.87$\pm$0.28
by PSF fitting. For the aperture photometry we used a smaller aperture with a 4 pixel
radius to avoid including flux from an image artifact to the south of the transient in 
the difference image (as can be seen in Fig. \ref{fig:subfig3}). Using the zeropoint
from aperture photometry (4 pixel radius) of the five SSCs, we find a magnitude
of $F222M$ = 17.59, with the same conservative error of $\pm$0.3 mag as estimated
for SN~2008iz. As a check on the effect of the small aperture on our photometry, we
used the same 4 pixel radius to measure the magnitude of SN~2008iz, and found a value
which differs only by 0.2 mag from that found through the 8 pixel aperture. As this 
difference is smaller than our errors, we do not regard this as a significant 
source of error.

From now on we adopt the PSF fitting based magnitudes as the most accurate ones to be
used in this study. We note that these are also consistent with the magnitudes obtained
through aperture photometry but have smaller uncertainties. As a final test we used
{\sc synphot} to calculate the $F222M-K$ colour of a range of black bodies between
100 and 50 000 K. For temperatures hotter than 500 K, we find that the colour
difference between the two filters is negligible (0.02 mag or less). While the
difference does become large ($>$0.1 mag) for T$<$400 K, if the flux in SN~2008iz
was coming from dust at this temperature, then it would be emitting in the mid-IR
rather than the $K$ band (\citeauthor{2008MNRAS.389..141M} 2008a).

\begin{figure}
\begin{center}
\includegraphics[width=60mm, angle=270, clip] {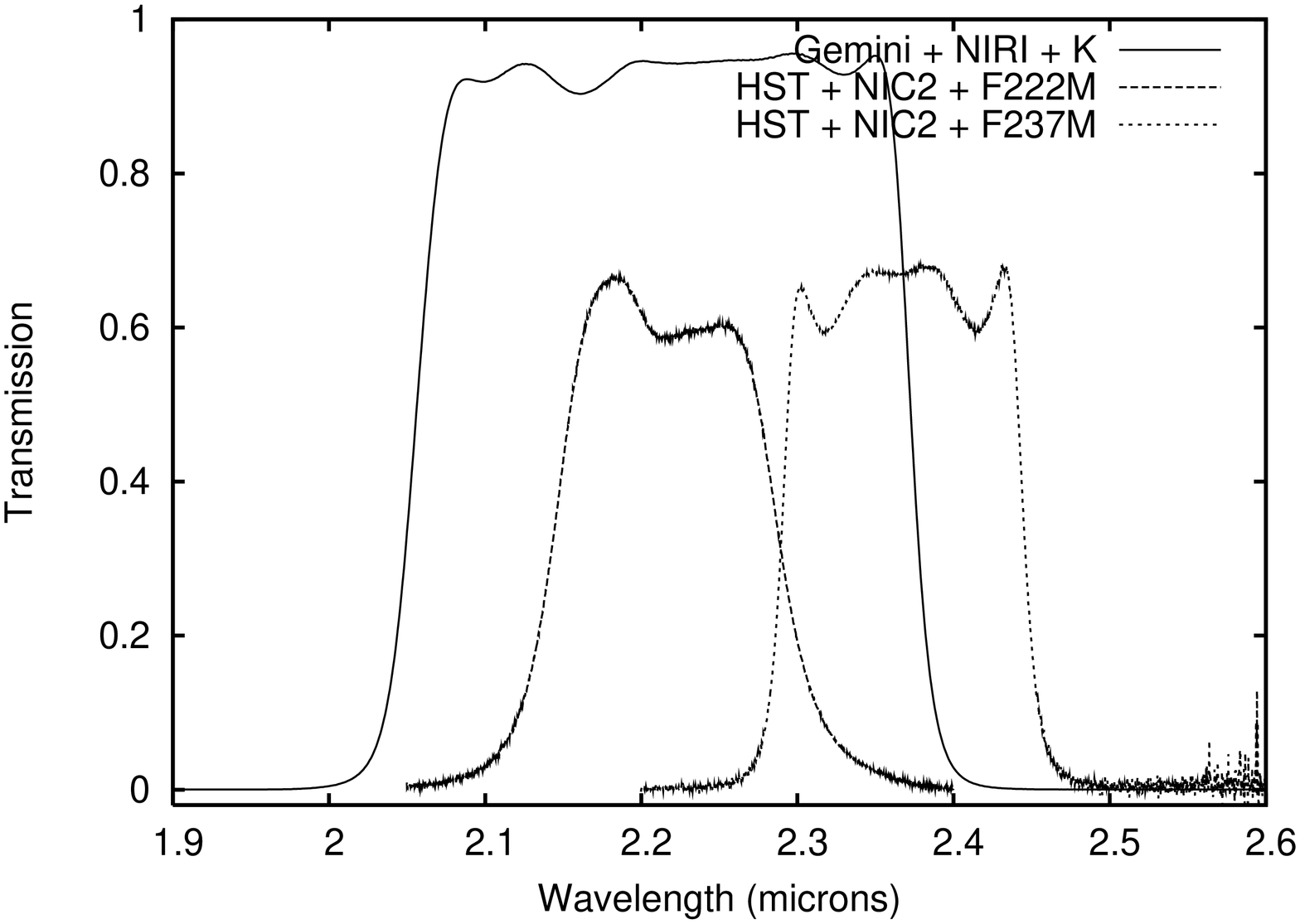}
\vspace{-0.3cm}
\caption{
Transmission curves for HST (telescope + instrument + filter) and Gemini (filter only)
used for the pre- and post-explosion imaging of SN~2008iz and the 43.78+59.3 transient.
}
\label{fig_filter}
\end{center}
\end{figure}

\section{Discussion}
\subsection{SN~2004am}
SN~2004am is the only SN ever discovered at optical/IR wavelengths in the 
prototypical starburst galaxy M~82. It occurred spatially coincident with the obscured SSC M~82-L and suffers 
from a host galaxy extinction of A$_{V}\simeq 5\pm1$ which is consistent with the extinction
derived for M~82-L.  From the compiled photometry and our NIR spectra it is
clearly a type II-P SN, which one would expect to come from a red supergiant
progenitor star.  It is likely a low luminosity type II-P SN and these have been proposed 
to come from stars toward the lower mass range that will produce core-collapse 
(Smartt et al. 2009). 

\citeauthor{2008A&A...486..165L} (2008) and \citeauthor{2007ApJ...663..844M} (2007) show that the NIR spectra
of M~82-L are dominated by the spectral features of red supergiant stars. The virial mass
of the cluster is estimated to be $4\pm0.6\times10{^6}$\msol\ 
(\citeauthor{2007ApJ...663..844M} 2007) which agrees with the mass estimate from the cluster luminosity by 
\citeauthor{2008A&A...486..165L} (2008). Assuming the best fit age of around 18\,Myr
(Sect.\ref{ssc-progenitor}) for M~82-L, the single stellar population models of \citeauthor{2008A&A...486..165L} (2008)
suggest there are likely to be of the order of 100-260 red supergiants in the cluster, 
depending on the initial mass function and lower mass limit used. Clearly there is a rich population of 
red supergiants as potential progenitors of type II-P SNe. 

In fact M~82-L  is the most massive host cluster of a SN that has been
accurately measured in the nearby Universe. Nearby SNe do not tend to
be discovered coincident with the most massive unresolved clusters. In the volume
limited sample of 20 II-P SNe reviewed by \citeauthor{2009MNRAS.395.1409S} (2009) which have
high-resolution pre-explosion images available, only two fall on compact clusters. 
This may be a little surprising, however \citeauthor{2007ApJ...658L..87P} (2007) show that in
NGC~1313 75-90 per cent of the UV flux is produced by stars outside the dense clusters.
They propose that the late O and early B-type stars (that will eventually produce
8-30 \msol\ red supergiants) are diffusely spread throughout a galaxy
due to infant mortality of stellar clusters (\citeauthor{2003ARA&A..41...57L} 2003).
\citeauthor{2008ApJ...672L..99C} (2008) have shown that the type Ic SN~2007gr 
exploded on the edge of a bright object in NGC 1058 and suggested that 
this may be a host cluster. This is one of 12 type Ib/c SNe studied
by Eldridge et al. (2013) at high resolution and is the only one possibly 
associated  with a stellar cluster, again supporting an approximate
figure of 10 per cent. 

The mass range that we suggest is appropriate for the progenitor 
($12^{+7}_{-3}$\,\msol) is in agreement with the masses of red supergiants 
directly detected in nearby galaxies. \citeauthor{2009MNRAS.395.1409S} (2009) present a 
full list of a volume and time limited search for SN progenitors, and find the 
likely progenitor population of II-P SNe are red supergiants of initial masses 
8-17~\msol. The mass of the progenitor of SN~2004am is consistent with this mass 
range. The fact that it was in a cluster with such a rich red supergiant 
population further supports the idea that this stellar evolutionary phase 
directly gives rise to hydrogen rich II-P SNe. However, we note that 
\citeauthor{Fra11} (2011) suggest a lower mass range of between 8-9 \msol ~for the 
progenitors of sub-luminous Type II-P SNe, and if SN~2004am was one of 
these events, it is also in agreement with the expected progenitor mass range
within the 1$\sigma$ uncertainty.

\subsection{SN~2008iz}
\begin{figure}
\begin{center}
\includegraphics[width=90mm, angle=0, clip] {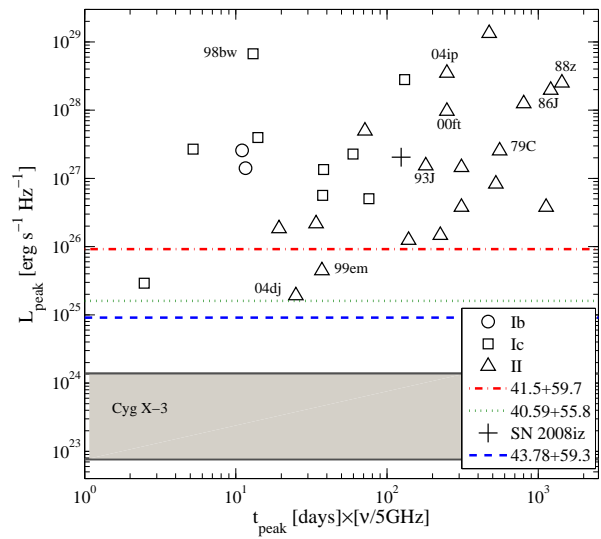}
\vspace{-0.3cm}
\caption{
Peak spectral radio luminosity versus time to reach the peak (normalized to
5\,GHz) for different types of CCSNe. The four M~82 transients have been added
to allow comparison with historic radio SNe. SN\,2008iz is represented by a
cross, whereas the rest of the transients, for which the peak epoch is unknown,
are represented as dash-dotted (41.5+59.7), dotted (40.59+55.8) and dashed
(43.78+59.3) lines. The gray region represents the range that the
galactic microquasar Cygnus X-3 has been observed to reach in major flare states
at radio wavelengths (\citeauthor{1995AJ....110..290W} 1995).}
\label{fig:Ltradio}
\end{center}
\end{figure}

In Fig. \ref{fig:Ltradio}, we compare the radio luminosities of SN~2008iz and the
two radio transients in M 82 (41.5+59.7 and 40.59+55.8) of unknown nature with the peak spectral
radio luminosities of SNe with different types. The luminosities of the two transients
are comparable with those of type II-P SNe, such as SNe 1999em and 2004dj. The peak luminosity
and time to reach the peak make SN~2008iz very similar to the well observed type IIb SN\,1993J
thus suggesting that also their progenitor stars might be similar. However, the expansion velocity
of $\sim$20 000\,km\,s$^{-1}$ observed for SN 2008iz (Brunthaler et al. 2010) was somewhat
higher than the one for SN 1993J ($\sim$15\,000\,km\,s$^{-1}$; \citeauthor{1995Sci...270.1475M} 1995).

At a distance of 3.3 Mpc, SN~2008iz has an absolute magnitude of K= -11.68$\pm$0.16 for
zero host galaxy extinction. The first detection of SN~2008iz at radio wavelengths was in March 2008,
with a best estimate for the explosion epoch of 2008 Feb. 18, implying that at the time of our
NIRI observation the SN was already $\sim$480 days old. Unfortunately there are few SNe with NIR
light curves for comparison at this late phase. However, SN~1987A had an absolute magnitude of
K= -12.2 (\citeauthor{1990AJ.....99..650S} 1990) at +450 days, while at 300 days SN~2004et had an absolute
magnitude of K = -13.35 (\citeauthor{2010MNRAS.404..981M} 2010). These magnitudes are comparable to that 
of SN~2008iz with a modest amount of extinction (A$_K <$ $\sim$1 mag).

\citeauthor{2010A&A...516A..27B} (2010) estimated the extinction towards SN~2008iz to be
$A_V$ = 24.4 mag, using the intensity of the $^{12}$CO $(J=2\rightarrow1)$ line 
from \citeauthor{Wei01} (2001), together with the CO to H$_2$ ratio and the relation of \citeauthor{Guv09} (2009)
between hydrogen column density and extinction. We find however, that when using 
the same data and relations as \citeauthor{2010A&A...516A..27B} (2010), we obtain a total
extinction of A$_V$ = 48.9 mag. This is exactly twice the value found by \citeauthor{2010A&A...516A..27B} (2010),
and we suggest that these authors may have neglected to multiply the value of $N(H_2)$ by two to
convert to hydrogen nuclei before applying the hydrogen nuclei to extinction relation of \citeauthor{Guv09} (2009).

A$_V$ = 48.9 mag implies an extinction in $K$ of $\sim$5.5 mag (\citeauthor{1989ApJ...345..245C} 1989).
Such a high extinction would put SN~2008iz at an absolute magnitude of $K$ $\sim$ -17 mag,
which is uncomfortably high for a SN at such a late phase. While a late-time near-infrared excess
has been observed for several SNe, and attributed to dust formation in the cool, dense
shell located between the forward and reverse shocks (e.g. \citeauthor{Fox09} 2009), a K-band 
excess of this magnitude is quite unusual. Furthermore, such excesses are
most commonly seen in Type IIn SNe, which are intrinsically rare. For example, adopting a
distance of 18.6 Mpc (assuming the Virgo+GA+Shapley corrected recession velocity from NED and
H$_{0}$ = 70 km s$^{-1}$ Mpc$^{-1}$) for the Type IIn SN~1998S its absolute $K$-band magnitude
(\citeauthor{2004MNRAS.352..457P} 2004) at an epoch of $\sim$480 days was $\sim$ -17.5. A more likely explanation
of the apparent discrepancy between the extinction for SN~2008iz as derived from the H$_2$ column density, and that
expected for a SN at that phase is that much of the H$_2$ lies behind the site of SN~2008iz. If the SN was located
behind one fifth of the total H$_{2}$ column, then its derived extinction would be closer to $A_V \sim$ 10 mag, bringing the 
late time magnitude into agreement with that of SNe 1987A and 2004et.

\subsection{The 43.78+59.3 transient}
One of the possible explanations advocated for the 43.78+59.3 transient by \citeauthor{2010MNRAS.404L.109M} (2010)
and \citeauthor{2011MNRAS.415L..59J} (2011)
was that it was an extra-galactic microquasar. Microquasars are the stellar analogs to quasars,
and are believed to arise from a compact object accreting matter from a stellar companion, and in
the process forming a relativistic jet. They typically display strong and variable radio and X-ray emission.
In Figure \ref{fig:Ltradio} we compare the radio luminosity of the 43.78+59.3 transient
with the range of luminosities that the Galactic microquasar Cygnus X-3 has been observed to reach
in major flare states (\citeauthor{1995AJ....110..290W} 1995). Only if Cygnus X-3 would increase its
highest observed radio luminosity by a factor of $\sim$10, the 43.78+59.3 transient would be comparable in
brightness. We note that recently \citeauthor{2012A&A...542L..24B} (2012) have proposed that three rapidly
variable radio sources within the
nuclear regions of the ultraluminous infrared galaxy Arp 220 having their radio luminosities a factor
$\sim$100 times larger than the luminosity of the 43.78+59.3 transient could be associated with highly beamed
microquasars. The apparent magnitude of the 43.78+59.3 transient is $K$ $\sim$ 17.9, corresponding to an absolute 
magnitude of $K$ $\sim$ -9.7 if assuming no host galaxy extinction. Such a bright absolute magnitude would make
the 43.78+59.3 transient the brightest known example of a microquasar by quite some margin; for comparison, the Galactic 
micro quasar SS433 has an absolute magnitude which varies between $K$=-5.7 and $K$=-6.4 
(\citeauthor{1985ApJ...296..232K} 1985; \citeauthor{2004ApJ...616L.159B} 2004). 

The 43.78+59.3 transient was first detected in the radio 40 days prior to the epoch of the Gemini 
image. From the templates presented in \citeauthor{2001MNRAS.324..325M} (2001) at +40 days, we 
would expect a normal, unextinguished SN in M~82 to have an apparent magnitude of 
$K$$\sim$9.6. Clearly the observed magnitude of the 43.78+59.3 transient appears difficult to
reconcile with a SN. If, however, the first detection of the 43.78+59.3 transient did not mark the explosion 
epoch, but was instead late time radio emission, then the faint absolute magnitude is less
puzzling. Similarly, $\sim$70 magnitudes of extinction in V would bring the observed and
expected $K$ magnitudes into agreement. However, even with a high extinction or an earlier
explosion epoch, its observed radio flux density evolution (\citeauthor{2012arXiv1209.6478G} 2012; Rob Beswick,
private communication) points to a non-SN origin for this source.

The absolute magnitude of the 43.78+59.3 transient is comparable to the quiescent magnitude 
of Luminous Blue Variables (LBV). LBV eruptions, also termed ``supernova impostors''
(\citeauthor{2000PASP..112.1532V} 2000) can reach an absolute magnitude of $\sim$ -12 to -16 in $V$. The 
SN impostors SN~2009ip and UGC 2773 OT2009-1 have upper limits on their 8 GHz 
radio luminosity of $<1.3 \times 10^{26}$erg s$^{-1}$ Hz$^{-1}$ and $<2.6\times 10^{26}$ erg 
s$^{-1}$ Hz$^{-1}$ respectively (\citeauthor{Fol11} 2011). These limits are an order of magnitude 
higher than the peak radio luminosity of the 43.78+59.3 transient, and hence the luminosity of the
latter is not inconsistent with the limited information on the properties of SN impostors at radio 
wavelengths. 

The NIR magnitude of the 43.78+59.3 transient is also reminiscent to that of a nova at peak. 
Most common novae fade rapidly after outburst, but from the template light curves
presented in \citeauthor{Str10} (2010) either an F or D-class nova could still be bright at $\sim$+40 days. 
However, classical novae are not usually strong radio sources, and the peak of the radio 
light curve tends to be later than the optical maximum, hence the MERLIN
detection would appear incongruous with this scenario.

The nature of the 43.78+59.3 transient remains elusive, and on the basis of the limited data available,
we regard an extremely bright extragalactic microquasar as the most plausible scenario, perhaps 
from a high-mass X-ray binary such as LS 5039 (\citeauthor{Cla01} 2001), but with a high ratio of radio to
X-ray flux. A bright extragalactic microquasar was proposed as the most likely explanation by
\citeauthor{2010MNRAS.404L.109M} (2010) and \citeauthor{2011MNRAS.415L..59J} (2011). We concur with that
conclusion although it would mean that the NIR luminosity is a factor of about 30 higher than Galactic microquasars.

\subsection{Comparison with the expected SN rate}
SNe 2004am and 2008iz are still the only supernovae detected at optical/NIR wavelengths in M~82.
This is perhaps somewhat surprising given the SN rate estimates of around 7-9 per century
(\citeauthor{2008MNRAS.391.1384F} 2008; \citeauthor{2010MNRAS.408..607F} 2010), the small distance and the
appeal of this galaxy as a
target in SN searches. It is very likely that extinction has hampered the discovery of recent SNe in the
past decade (for estimates of extinctions towards the radio SNRs of M~82 see \citeauthor{2001MNRAS.324..325M} 2001).
Over the last $\sim$20-30 years during which M~82 has been regularly monitored at radio wavelengths,
the detection of two confirmed SNe (SNe 2004am and 2008iz) and the two radio transients (41.5+59.7 and 40.59+55.8) with
a possible SN origin is in reasonable agreement with the expectation. However, the discoveries of SNe
2004am and 2008iz have important lessons for attempts to find heavily extinguished SNe in
more distant starburst galaxies. While SN~2008iz was recovered at a comparable magnitude 
to the two SNe found by \citeauthor{2012ApJ...744L..19K} (2012) in their NIR LIRG SN search, it is of note
that \citeauthor{Fra09} (2009) originally reported a non-detection for both SN~2008iz and the 43.78+59.3 transient. 
We have only recovered these transients with a careful analysis using {\it a posteriori} 
knowledge of the source positions from radio observations. Without these data, it is likely that 
SN~2008iz would have remained undiscovered in the NIR. 

Recently \citeauthor{2012arXiv1206.1314M} (2012)
studied the fraction of CCSNe missed by rest-frame optical SN searches and found an average local
value of $\sim$20\% increasing to $\sim$40\% of CCSNe missed at z $\sim$1-2. These estimates
highlight the need for a better control of the SN activity in the obscured environments such
as the nuclear regions of nearby starburst galaxies and LIRGs. Unless properly accounted
for such systematic effects can dominate the uncertainties in the CCSN rates at high redshift
(e.g., \citeauthor{melinder2012} 2012; \citeauthor{2012ApJ...757...70D} 2012).

\section{Conclusions}
SN~2004am was the first optically detected SN in M~82 when discovered by LOSS (Singer et al. 2004),
and remains the only transient discovered in the optical. We show that it is most likely to have been
a sub-luminous, highly reddened Type II-P event coincident with the obscured nuclear SSC M~82-L. From the
cluster age we infered a progenitor mass of $12^{+7}_{-3}$\,\msol which is within the uncertainties in agreement
with the expected progenitor mass range for such events.
Making use of high spatial-resolution $K$-band imaging we detected NIR counterparts for both
SN~2008iz and the 43.78+59.3 transient previously detected only at radio wavelengths. Our late-time $K$-band magnitude
rules-out an extremely high extinction towards SN~2008iz. The nature of the 43.78+59.3 transient still remains 
elusive, an extremely bright microquasar in M~82 being the most plausible scenario.

\section*{Acknowledgments}

We thank an anonymous referee for several useful suggestions, Ariane Lan{\c c}on for providing the NIR spectrum
of M~82-L, Catherine Walsh for helpful discussions on extinction and atomic and molecular column densities, Rubina Kotak
for carrying-out some of the WHT observations, Andrea Pastorello for helpful discussions, and Rob Beswick for
providing information on the 43.78+59.3 transient. The research leading to these results has received
funding from the European Research Council under the European
Union's Seventh Framework Programme (FP7/2007-2013) ERC Grant
agreement n$^{\rm o}$ [291222] (PI : S. J. Smartt). S. Mattila
and C. Romero-Ca\~nizales acknowledge financial support from the
Academy of Finland (project: 8120503) and M. Fraser acknowledges
support from the UK's STFC.

Based on observations made with the Nordic Optical Telescope, operated
on the island of La Palma jointly by Denmark, Finland, Iceland,
Norway, and Sweden, in the Spanish Observatorio del Roque de los
Muchachos of the Instituto de Astrofisica de Canarias. 
The William Herschel Telescope is operated on the island of La Palma by the Isaac 
Newton Group in the Spanish Observatorio del Roque de los Muchachos 
of the Instituto de Astrof'sica de Canarias. 

Based on observations obtained as part of program GN-2009A-Q-32 at the Gemini 
Observatory, which is operated by the Association of Universities for Research 
in Astronomy, Inc., under a cooperative agreement with the NSF on behalf of 
the Gemini partnership. Also based on observations made with the NASA/ESA 
Hubble Space Telescope, obtained from the Data Archive at the Space 
Telescope Science Institute, which is operated by the Association of 
Universities for Research in Astronomy, Inc., under NASA contract NAS 5-26555.

\end{document}